\RequirePackage{fix-cm}
\documentclass[twocolumn]{svjour3}  
\smartqed  
\RequirePackage{graphicx}
\RequirePackage{mathptmx}      
\RequirePackage{latexsym}
\RequirePackage[numbers,sort&compress]{natbib}
\RequirePackage[colorlinks,citecolor=blue,urlcolor=blue,linkcolor=blue]{hyperref}
\usepackage{newtxtext}    
\usepackage{newtxmath}    
\usepackage{microtype}    
\usepackage{xspace}
\usepackage{graphicx}
\usepackage{url}
\usepackage{xurl}
\usepackage{amsmath}
\usepackage{booktabs}
\usepackage{makecell}
\usepackage{tcolorbox}
\usepackage{adjustbox}
\usepackage{tikz}
\usepackage{footmisc}
\usepackage{balance}
\usepackage{titlesec}

\renewcommand{\note}[1]{#1\xspace}

\newcommand{\eg}{\emph{e.g.\xspace}}

\newtcolorbox{cbox}[1]{
    colback=gray!10,
    colframe=black,
    title=#1,
    fonttitle=\footnotesize,
    fontupper=\footnotesize,
    before skip=10pt,    
    after skip=0pt,     
}

\newcommand{\tightbox}[1]{%
  \tikz[baseline=(X.base)] \node[%
    draw=black,           
    fill=gray!10,         
    rounded corners=2pt,  
    inner xsep=2pt,       
    inner ysep=2pt,       
    anchor=base
  ] (X) {#1};%
}

\makeatletter
\renewcommand{\@maketitle}{%
  \newpage
  \null
  \vskip 2em%
  \begin{center}%
    {\LARGE \@title \par}%
    \vskip 1.5em%
    {\large
      \lineskip .5em%
      \begin{tabular}[t]{c}%
        \@author
      \end{tabular}\par}%
    \vskip 1em%
    {\@date}%
  \end{center}%
  \par
  \vskip 1.5em}
\makeatother
\journalname{}

\begin{document}

\title{Analysis of Personal Data Exposure in Thailand\thanks{Accepted for publication in the \href{https://link.springer.com/journal/10207}{International Journal of Information Security}, April 30, 2026.}
}



\author{%
    Suphannee Sivakorn\textsuperscript{1}
    \and
    Sasawat Malaivongs\textsuperscript{2}
    \and
    Nuttaya Rujiratanapat\textsuperscript{1}%
}





\institute{Suphannee Sivakorn \at \email{suphannee\_si@rmutto.ac.th}
    \and
    Sasawat Malaivongs \at \email{sasawat@acinfotec.com}
    \and
    Nuttaya Rujiratanapat \at \email{nuttaya\_ru@rmutto.ac.th}
    \and
    \textsuperscript{1}Department of Computer Science, 
    Faculty of Science and Technology\\
    Rajamangala University of Technology Tawan-ok, Thailand
    \\\textsuperscript{2}ACinfotec Co., Ltd., Thailand
}

\date{}

\maketitle


\begin{abstract}
In the digital era, personal data, particularly sensitive identifiers such as the Social Security Number and National Identification Number, has become a highly valuable asset, raising significant concerns regarding privacy and security. This study examines the risks associated with the online exposure of the \emph{Thai National Identification Number}, a key element of identity verification in both governmental and commercial transactions. Similar to the Social Security Number in the United States, this unique identifier is crucial for various legal, financial, and welfare-related activities.  However, the increasing digitization of personal records has heightened its vulnerability to unauthorized access and misuse, particularly through search engines that inadvertently index sensitive information.  

This research identifies publicly exposed Thai National Identification Numbers across major search engines, assessing the potential threats to individual privacy and national security. The study reveals the exposure of over 1.2 million unique National Identification Numbers, along with other highly sensitive personal data, e.g., addresses, contact details, employment status, disability status, and health information. Notably, the analysis indicates that a significant majority of these exposures originate from the Thai government sector websites, highlighting critical vulnerabilities in public data management practices. This widespread exposure not only increases the risk of identity theft and financial fraud but also underscores the urgent need for enhanced cybersecurity measures, stricter regulatory enforcement, and improved data governance within government agencies to prevent future breaches. Addressing these issues is essential to safeguarding citizens' personal information and ensuring compliance with Thailand's data protection laws in an increasingly digitized world.
\end{abstract}

\section{Introduction}
\label{sec:intro}

In the digital age, personal data has become a commodity of immense value, facilitating transactions, services, and often serving as a gateway to individual identities. Thailand, like many other nations, faces significant challenges in safeguarding the privacy and security of its citizens' personal information, particularly concerning the Thai National Identification Number. Similar in function to Social Security Number (SSN) in the United States, the Thai National Identification Number is issued by the Thai government to individuals who are Thai citizens and/or registered in the household registration system. The Identification Number serves as evidence of identity, proof of personal status, and confirmation of identity for government transactions, service requests, and welfare benefits from state agencies. It is also used for various commercial transactions, legal activities, and other purposes such as job applications, opening bank accounts, asset and property transfers. However, its misuse and exposure through online platforms pose substantial threats to both individual privacy and national security.

The proliferation of Internet usage and the digitization of personal records have inadvertently exposed sensitive information, including Thai National Identification Numbers, to online platforms. Search engines, designed to index and retrieve vast amounts of information, inadvertently become repositories of personal data. This exposure creates fertile ground for malicious actors engaged in identity theft, financial fraud, and other forms of cybercrime~\cite{predescu2023implications,clough2011data}. The ease with which such information can be harvested and exploited underscores the urgent need for enhanced cybersecurity measures and regulatory frameworks.

Existing legal frameworks, such as Thailand's Personal Data Protection Act (PDPA) B.E. 2562 (2019)~\cite{thaipdpa}, aim to regulate the collection and use of personal data. However, the widespread exposure of sensitive information online suggests gaps in enforcement and the implementation of cybersecurity best practices. Previous studies on personal data exposure and cybercrime have focused on similar issues in other countries~\cite{zulfiani2021prevention,vijay2024survey,SADHYA2024103782,10.1145/1592665.1592668,holtfreter2015data}, but research specific to the exposure of Thai National Identification Numbers remains scarce.

In an effort to understand the prevalence and implications of personal data exposure online, this study investigates the current state of personal data leaks as revealed through search engine results. Utilizing a systematic search methodology across major search engines, we identify and collect search results related to personal information, with a particular focus on the Thai National Identification Number. These results are then analyzed across multiple dimensions to identify key patterns, assess the primary root causes of data exposure in Thailand, and propose strategies to mitigate future leaks and prevent the malicious exploitation of exposed data.

The key contributions of this study are as follows:
\begin{itemize}
    \item To the best of our knowledge, this is the first practical study to systematically analyze personal data exposure online in Thailand. This research evaluates the prevalence and underlying factors contributing to personal data exposure, offering a data-driven perspective on the issue.
    
    \item We designed and implemented an automated personal data collection system capable of retrieving and extracting exposed personal information, particularly Thai National Identification Numbers from online sources. This system serves as a monitoring tool for identifying personal data leaks and safeguarding sensitive information. Through this approach, we successfully collected over 1.2 million instances of exposed personal data.

    \item Our analysis identifies the primary sources of personal data leak, revealing that a significant proportion originates from government sector websites, particularly local government offices. Additionally, we uncover specific data patterns and instances of publicly disclosed sensitive information that could be exploited for malicious activities such as phishing, scams, and identity theft, posing a direct threat to individual privacy.
\end{itemize}

The remainder of this paper is structured as follows: Section~\ref{sec:background} provides background information and discusses related work, including personal data regulations, relevant laws, and the theoretical context surrounding the Thai National Identification Number, personal data exposure, and cybercrime trends in Thailand. Section~\ref{sec:tech} presents the technical background, covering search engine indexing, website structures, domain names, and Thailand's domain name system. Section~\ref{sec:method} details our methodology for developing the personal data collection system, including keyword selection, data retrieval processes, and personal data extraction techniques. Section~\ref{sec:eval} presents the analysis of the collected data, including statistical insights and key findings. We explore and discuss general countermeasures against the personal data exposure in Section~\ref{sec:counter}. 
Section~\ref{sec:ethics} discusses the ethical considerations of this research, including responsible disclosure practices. 
Section~\ref{sec:related} reviews the related work.
Finally, Section~\ref{sec:conclusion} concludes the paper and outlines future research directions.

\section{Background}
\label{sec:background}

\subsection{Personal Data and Thai Personal Data Protection Act B.E. 2562 (2019)}
\subsubsection{Overview}
The Thai Personal Data Protection Act (PDPA), B.E. 2562 (2019), Thailand's first comprehensive data protection law, was published in the Royal Thai Government Gazette (Ratchakitchanubeksa) on May 27, 2019, and became fully effective on June 1, 2022. Modeled closely on the European Union's General Data Protection Regulation (GDPR)~\cite{gdpr}, the PDPA seeks to align with global data privacy and protection standards. By adopting many of the GDPR’s principles, the law aims to strengthen data privacy in Thailand and ensure consistency with international best practices.

Like the GDPR, this law grants individuals significant control over their personal data, ensuring that the data collection, processing and storage activities are transparent, lawful and secure. Both regulations emphasize the importance of obtaining explicit consent from data subjects and provide individuals with a range of rights over their data, such as the right to access, correct, and delete personal data, as well as the right to withdraw consent and transfer data across borders. In terms of enforcement, the PDPA also follows the GDPR's approach by establishing strict penalties for non-compliance and for breaches of data protection regulations.

However, when it comes to enforcement, there is a notable difference in the penalty structures. The GDPR allows for fines based on a percentage of an organization's annual global revenue up to 4\% of annual global turnover or 20 million EUR, whichever is higher. In contrast, the PDPA imposes fixed penalties, with fines not exceeding 5 million THB (approximately 141,000 USD)~\cite{gdpr_pdpa}.
Both the PDPA and the GDPR emphasize the protection of sensitive personal data (such as health, religious beliefs, and biometric information), recognizing the increased risk of harm that can result from unauthorized disclosure or misuse of such information.

Although the PDPA is tailored to the context of Thailand's legal and cultural framework, it shares many similarities with the GDPR, positioning Thailand as part of the global movement toward stronger data protection practices.

\subsubsection{Key Categorizations and Personal Data Owner Rights}
The PDPA defines ``Personal Data" and ``Sensitive Personal Data" as follows:

\textbf{Personal Data} refers to any information related to an individual that can be used to identify that person, either directly or indirectly. Examples of personal data include names, 
various identification numbers (\eg Thai National Identification Numbers, passport numbers, bank account numbers, credit card numbers), contact information (\eg addresses, email addresses, phone numbers), and asset and property information (\eg vehicle registration numbers, house registration number), and other data that can be linked to personal information (\eg date and place of birth, 
medical records, educational records, financial information, employment records). 

\textbf{Sensitive Personal Data} includes high-risk information such as race, ethnic origin, political opinions, cult, religious or philosophical beliefs, sexual behavior, criminal records, health information, disabilities, 
biometric data or of any data which may affect the data subject in the same manner.  
%
The PDPA imposes stricter safeguards and higher penalties for the unauthorized disclosure of these categories due to their potential for discriminatory misuse.


\textbf{Personal Data Owner Rights.} Under the PDPA, data owners are entitled to access and obtain copies of their personal data, request corrections, and demand erasure or destruction when appropriate. They may withdraw consent at any time, request data portability for information obtained directly from them, and object to processing undertaken on legal grounds. Additionally, they may request temporary restrictions on data processing. Organizations must comply 
or provide documented justification for any refusal.

\subsection{Thai National Identification Number}
\label{sec:thai_id_num}
The \emph{Thai National Identification (ID) Number} or \emph{Thai ID Number} or \emph{Citizen ID Number}, is a 13-digit number displayed on the Thai National ID Card, a government-issued document for Thai citizens who are registered in the household registration system. This card serves as a primary form of identification for Thai nationals and is used to verify and authenticate an individual's identity.
%
The National ID number serves as the primary key for accessing services, financial systems, and legal transactions in Thailand, particularly for digital authentication and verification purposes.

\subsubsection{The Meaning of National ID Number}
The Thai National ID Number, consisting of 13 digits, holds specific meanings for each digit, as outlined below.

\begin{figure}[b]
    \centering
    \includegraphics[width=0.8\columnwidth]{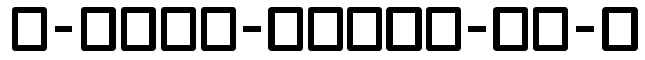}
    \caption{Format of National ID Number as Displayed on the Thai National ID Card}
    \label{fig:cid_format}
\end{figure}

\textbf{Digit 1 (Individual Classification):} signifies the category of the individual. These categories, detailed in Table~\ref{tab:id_categories}, are fundamental to the demographic analysis of data exposure patterns conducted in this study.

\begin{table}[h]
\caption{Classification of Individuals by the First Digit of the National ID Number}
\label{tab:id_categories}
\centering
\footnotesize
\begin{tabular}{|c|p{6.5cm}|}
\hline
\textbf{Number} & \textbf{Description of Individual Category} \\ \hline
1 \& 2 & Thai nationals born after Jan 1, 1984, with the number 1 indicating those who registered their birth within 15 days, and the number 2 indicating those who failed to register within that period. \\ \hline
3 & Thai nationals/foreigners in the household registry prior to May 31, 1984. \\ \hline
4 & Thai nationals/foreigners establishing residency before an initial ID assignment. \\ \hline
5 & Individuals added later due to previous omissions or special circumstances. \\ \hline
6 & Temporary residents, illegal entrants, or ethnic groups awaiting citizenship. \\ \hline
7 & Children born in Thailand to parents classified under Category 6. \\ \hline
8 & Legal foreign residents or those acquiring Thai nationality after May 31, 1984. \\ \hline
\end{tabular}
\end{table}



\textbf{Geographic Identifiers (Digits 2--5):} represents the issuing registration office: Digit 2 denotes the administrative region (1--9), Digit 3 identifies the province, and Digits 4--5 specify the district or municipality.



\textbf{Sequential Identifiers (Digits 6--12):} serve as internal classification groups or sequential birth certificate numbers, providing a unique serial identifier within the local registry.

\textbf{Checksum Validation (Digit 13):} is the final checksum digit, calculated to verify the correctness of the preceding 12 digits of the National ID number.

\subsubsection{National ID Number Checksum Validation}
\label{sec:thai_id_valid}
Here, we describe the method used to calculate the checksum (Digit 13). For malicious actors, this checksum facilitates ``data scrubbing," allowing any leaked numeric string to be programmatically verified as a valid National ID, thereby increasing the efficiency of mass data harvesting.

\textbf{Step 1:} Multiply each digit of the National ID number by its corresponding \emph{position multiplier} (from 13 down to 2), then sum all the results. Table~\ref{tab:cid_checksum} presents the example of calculating the sum of all multiplication results (351)


\begin{table*}[ht]
\centering
\caption{Example of Checksum Calculation for the 13th Digit Verification}
\label{tab:cid_checksum}
\small
\begin{tabular}{|l|c|c|c|c|c|c|c|c|c|c|c|c|c|}
\hline
\textbf{Position} & \textbf{1} & \textbf{2} & \textbf{3} & \textbf{4} & \textbf{5} & \textbf{6} & \textbf{7} & \textbf{8} & \textbf{9} & \textbf{10} & \textbf{11} & \textbf{12} & \textbf{\begin{tabular}[c]{@{}c@{}}13\\ (Checksum)\end{tabular}} \\
\hline
\textbf{Position Multiplier} & 13 & 12 & 11 & 10 & 9 & 8 & 7 & 6 & 5 & 4 & 3 & 2 & \multicolumn{1}{l|}{} \\
\hline
\textit{\textbf{Example}} & \textit{\textbf{1}} & \textit{\textbf{2}} & \textit{\textbf{3}} & \textit{\textbf{4}} & \textit{\textbf{5}} & \textit{\textbf{6}} & \textit{\textbf{7}} & \textit{\textbf{8}} & \textit{\textbf{9}} & \textit{\textbf{1}} & \textit{\textbf{0}} & \textit{\textbf{1}} & \textit{\textbf{1}} \\
\hline
\textbf{Multiplication result} & 13 & 24 & 33 & 40 & 45 & 48 & 49 & 48 & 45 & 4 & 0 & 2 &  \\
\hline
\multicolumn{1}{|c|}{\textbf{Checksum Validation}} & \multicolumn{12}{c|}{351} & 
    $ 11 - (351 \mod 11) = \emph{\textbf{1}} $ \\
\hline
\end{tabular}
\end{table*}

\textbf{Step 2:} Divide the result obtained in Step 1 by 11 and calculate the remainder (modulus operation):
\begin{center}
$ 351 \mod 11 = 10 \nonumber $
\end{center}

\textbf{Step 3:} Subtract the remainder obtained in Step 2 from 11. In this example:
\begin{center}
$ 11 - 10 = 1 \nonumber $
\end{center}

\textbf{Step 4:} If the result of the subtraction is a two-digit number, use the units digit. Therefore, the 13th digit of the ID number is 1.

\subsection{Cybercrime Landscape in Thailand}
\label{sec:cybercrime_data}

Cybercrime in Thailand has been on the rise in recent years. According to statistics from the Royal Thai Police, between March 2022 and May 2023, over 296,000 cybercrime-related reports were filed, an average of 525 cases per day, resulting in financial damages exceeding 40 billion THB (approximately 1.2 billion USD), or an average of 74 million THB per day. The majority of these crimes involve various forms of online scams and fraud
which include Fraudulent online transactions involving goods and services (37\%), Money transfer scams (13\%), Loan fraud (12\%), Investment scams (8\%) and Cyber extortion (7\%)~\cite{cybercrime_th}. 
This trend is driven by the widespread adoption of real-time payment systems, which, while efficient, have positioned Thailand sixth globally in online scam frequency~\cite{bot_money}. 


\subsubsection{Systemic Data Breaches and Identity Risk}
The efficacy of these scams is directly linked to the availability of leaked PII, which facilitates high-fidelity spear-phishing. Thailand has faced several landmark breaches that underscore this systemic vulnerability:

\textbf{Mass Population Exposure:} In 2024, a hacker known as ``9near" claimed 
the exposure of 55 million citizens' data, affecting nearly 83\% of the population~\cite{9near}.

\textbf{Commercial and Institutional Leaks:} Also in 2024, significant breaches involving 5 million records from loyalty programs~\cite{the_one_card} and the illicit sale of customer data by bank employees~\cite{emp_sell_data} highlight vulnerabilities across both private and financial sectors.

Beyond financial loss, these breaches inflict long-term \emph{identity scarring}, where victims face years of credit recovery and psychological distress~\cite{call_center1, call_center2}.

\subsubsection{Thai Cybercrime Law B.E. 2566 (2023)}
Recognizing the urgent need for a coordinated response to cybercrime, the Thai government enacted the Emergency Decree on Measures for the Prevention and Suppression of Technological Crimes, B.E. 2566 (2023)~\cite{thaicybercrime}. This legislation enhances collaboration between key stakeholders, including banks, telecommunications providers, internet service providers (ISPs), law enforcement, and financial institutions~\cite{lawplusltd}.
Key provisions of the decree include:
\begin{itemize}
    \item Strengthening information-sharing mechanisms among relevant authorities to detect and respond to cybercrime.
    \item Granting account holders the right to temporarily freeze and report suspicious transactions to prevent fraud.
    \item Imposing stricter legal penalties on individuals involved in cybercriminal activities, such as, \emph{money laundering}, \emph{money muling}~\cite{money_mules}, and \emph{fraudulent identity schemes} \eg, opening bank accounts, registering phone numbers or creating fake social media profiles to impersonate government authorities for scam operations.
\end{itemize}


\subsection{Criminal Exploitation of National ID Numbers}
The exploitation of personal data, particularly the Thai National ID number, has become a critical concern in the digital age. Criminals leverage stolen personal information to engage in fraudulent activities, financial crimes, and identity theft, posing serious risks to individuals and institutions. 
Below are key examples of how the Thai National ID number can be misused for illicit purposes:

\textbf{Identity Theft and Financial Fraud.} Cybercriminals can exploit an individual's National ID number to impersonate them and gain unauthorized access to financial services~\cite{thaipbs_idcard}.
The risk is further heightened when National ID numbers are coupled with the Laser Codes, an alphanumeric identifier located on the back of the Thai National ID card, 
which allows criminals to bypass GDX-based authentication to open ``mule" accounts, apply for fraudulent loans, or intercept social welfare benefits~\cite{thai_gdx}.

\textbf{Online Phishing Scams.} Cybercriminals frequently use stolen personal data, including National ID numbers, in phishing schemes designed to deceive victims into divulging additional sensitive information. 
The more information they possess about a target, the more convincing and effective their phishing attacks become~\cite{alkhalil2021phishing}.


\textbf{Dark Web Monetization:} Verified citizen registries including National ID numbers are trafficked on hacker forums, where they serve as foundational datasets for long-term identity theft and automated scam operations~\cite{9near, emp_sell_data}.



\subsection{Emerging Risks: AI Training and Persistent Privacy Risks}
The risk of exposure of personal information has evolved considerably with the rapid proliferation of Large Language Models (LLMs). Large-scale web scraping of publicly accessible internet content constitutes a primary data source for training AI models~\cite{amarikwa2023internet,gogl_ai_bard}. This practice raises significant privacy concerns, as training pipelines frequently ingest vast amounts of public data without reliably filtering PII or obtaining explicit consent from data subjects~\cite{chatgpt_email,10253502,nyt_openai_sue}.
Such large-scale data ingestion introduces a persistent \emph{identity memorization} risk, whereby sensitive information, including National ID Numbers, may become implicitly embedded within model parameters. Unlike traditional databases, where compromised records can be deleted or access revoked, removing specific data instances from a trained model (\eg, machine unlearning) remains technically challenging and often infeasible in practice~\cite{ml_unlearn, ai_forget}. Consequently, the exposure of Thai citizens' personal data may result in long-term and potentially irreversible security and privacy risks within the global AI ecosystem.
\section{Technical Background}
\label{sec:tech}
This section presents the technical background necessary to understand the approach used in this study.

\subsection{Search Engine}
\label{sec:searchengine}
A search engine is a program designed as a tool for retrieving information from various online sources. Most search engines enable users to search for information on websites by entering keywords or phrases. The engine then processes the search query and presents relevant results to the user. The term entered for a search is referred to as ``search keyword".

\subsubsection{Search Engine Processes}
In general, search engines operate through several key stages to create and manage their databases. This section details the critical steps involved in search engine operations, which are essential for understanding the focus of this paper.

\begin{enumerate}
    \item \textbf{Crawling} refers to the process of collecting data from websites across the internet. Search engines gather information by following links within the content of a website that direct users to other sites. 
    \item \textbf{Indexing} involves creating an index of the content collected during the crawling process. This index is generated and stored based on the content of each website. While search engine indexes can be generated automatically by the search engine itself, website creators can customize their indexes using metadata (\eg,~meta tags)~\cite{gogl_crawling}, \texttt{robots.txt} files~\cite{gogl_crawling}, or other mechanisms. The primary objective of indexing is to organize the collected data in a way that allows it to be efficiently queried and used to deliver relevant search results in response to the user.
    \item \textbf{Ranking} is the process by which search results are ordered before being presented to users. The ranking of search results is determined by several factors, including the relevance of the search term to the indexed content, the quality of the content, and the volume of web traffic, among others. Typically, web pages that are most closely related to the search keyword are ranked higher, with results presented in descending order of relevance. This ensures that users can more easily access the information they are seeking.
\end{enumerate}

\subsubsection{Advanced Search}
In addition to basic keyword searches, modern search engines support advanced search features or special search operators, which enhance the accuracy and precision of search results~\cite{gogl_adv_search,ms_adv_search,ms_adv_search_keyword}.

\textbf{Examples of Advanced Search Techniques}
\begin{itemize}
    \item Quote (\texttt{" "}): This method involves using quotation marks around a search term to instruct the search engine to only return search results that contain the exact phrase. For example, entering \texttt{"dog"} in the search engine will return results that only contain the word ``dog" in the content~\cite{gogl_adv_search}.
    \item Exclude (\texttt{-}): This method involves using a minus sign before a search term to exclude results that contain that term. For example, \texttt{"-cat"} will show search results that do not contain the word ``cat".
    \item Filetype: This operator is used to search for specific file types. For example, using filetype:pdf will show results only from PDFs, and \texttt{filetype:doc} will return search results from DOC files~\cite{gogl_adv_search,ms_adv_search}.
    \item Site: This search operator restricts results to a specific website. For instance, \texttt{site:example.com} will return only search results from pages within the domain example.com ~\cite{gogl_adv_search}.
    \item Operators (\texttt{AND}, \texttt{OR}): This allows the user to refine their search based on specific conditions. For example:
    \begin{itemize}
        \item AND: Refines the search to return search results containing all specified terms. For example, \texttt{cat AND dog} will show search results that contain both ``cat" and ``dog."
        \item OR: Expands the search to return search results containing any of the specified terms. For example, \texttt{dog OR cat OR bird} will return search results containing any of the three terms.

    \end{itemize}
    \item Parentheses: This operator is used to group terms and organize the logic of the search, similar to its use in mathematics. For example, \texttt{dog AND (cat OR bird)} will show search results that contain the word ``dog" and either ``cat" or ``bird."
 
\end{itemize}

Users can combine these advanced search operators to refine their searches. For example:
\begin{itemize}
    \item \texttt{site:example.com filetype:pdf "dog" AND\\("cat" OR "bird")}
\end{itemize}

This query would return results with the following conditions:
\begin{itemize}
    \item Search results are limited to web pages within example.com (\eg~\texttt{www.example.com/pet}).
    \item The content must come from PDF files.
    \item The page must contain the word ``dog" and also either ``cat" or ``bird."
\end{itemize}

\subsection{URL and Domain Name}
\textbf{URL} (Uniform Resource Locator) is the address used to access a specific resource on the internet, such as a web page, image, or file. A URL serves as a standardized reference that allows users to locate and retrieve web content efficiently. The URL is composed of several key components, including the protocol, domain name, and path. For example, in the URL \texttt{https://example.com/page1}, \texttt{https} is the protocol, \texttt{example.com} is the domain name, and \texttt{/page1} is the specific path to the resource.

\textbf{Domain name} is a human-readable identifier used to represent an IP address in a more memorable and user-friendly format. Each domain name corresponds to an IP address, which is a numerical identifier used by computers to locate resources on the internet~\cite{rfc1035}. 
Domain names are used to identify websites, email servers, and other online services. 


\subsection{Top-Level Domain Name (TLD)}
Top-Level Domain (TLD) refers to the final segment of a domain name, which comes after the last dot. 
The TLD serves as a classification system for domain names and can denote various organizational, geographical or functional attributes.
For example:
\begin{itemize}
\item Generic Top-Level Domains (gTLDs): These are the most commonly used TLDs and consist of three or more alphabetical characters. Examples include \texttt{.com} (commercial), \texttt{.info} (information), and \texttt{.org} (organization). These TLDs are often used by a wide variety of organizations and businesses for general Internet purposes.

\item Country Code Top-Level Domains (ccTLDs): These TLDs are two-letter domain suffixes that are designated to represent specific countries or territories, according to the ISO 3166 international standard \cite{iso3166}. Examples include .us (United States), \texttt{.th} (Thailand), and \texttt{.cn} (China). ccTLDs are typically used to indicate a geographic or national connection and often serve as an indicator of the website's primary audience or origin.
\end{itemize}

\subsubsection{Second-Level Domain Name (SLD)}
The Second-Level Domain (SLD) resides immediately preceding the Top-Level Domain (TLD). It functions as a unique identifier within a broader domain namespace, potentially reflecting a specific category or purpose. While some TLDs, such as \texttt{.com}, stand alone without an SLD, others, like country code TLDs (\texttt{.th} in \texttt{"example.co.th"}), incorporate an SLD within their structure. This tiered system facilitates the organization and categorization of domain names based on their intended use or the entity they represent (\eg~government, education, businesses, non-profits). This hierarchical structure enhances user recognition of website ownership and purpose, contributing to a more streamlined and navigable domain name system. The subsequent section will delve into the specific categories of Thailand's SLDs.

\subsection{Thailand's Domain Name System and the \texttt{.th} ccTLD}
Thailand began its connection to the internet in 1986~\cite{thnic}.
Subsequently, the use of the \texttt{.th} Top-Level Domain (TLD) was introduced to represent domain names associated with Thailand. Today, the registration of domain names under the \texttt{.th} TLD is organized into seven categories of SLDs outlined in the Table~\ref{tab:th_tld}.
Each SLD category is intended to serve a specific sector, ensuring that domain registration aligns with the intended use, for example, governmental, academic, commercial, military, non-profit, individual, or network-related. 

\begin{table*}[ht]
\caption{Types and Purposes of SLD under the .th TLD}
\label{tab:th_tld}
\footnotesize
\setlength{\tabcolsep}{3pt}
\centering
\begin{tabular}{|l|p{5.3cm}|p{10cm}|}
\hline
\multicolumn{1}{|c|}{\textbf{\texttt{.th} TLD}} & \multicolumn{1}{c|}{\textbf{Purpose of Domain Name}} & \multicolumn{1}{c|}{\textbf{Example Domain Names}} \\ \hline
\textbf{\texttt{.go.th}} & Government agencies and organizations & \begin{tabular}[c]{@{}l@{}}\texttt{thaigov.go.th} (Thai Government), \\ \texttt{dla.go.th} (Department of Local Administration)\end{tabular} \\ \hline
\textbf{\texttt{.ac.th}} & Academic institutions and organizations & \begin{tabular}[c]{@{}l@{}}\texttt{rmutto.ac.th} (Rajamangala University of Technology Tawan-Ok), \\ \texttt{mahidol.ac.th} (Mahidol University)\end{tabular} \\ \hline
\textbf{\texttt{.co.th}} & Commercial entities and businesses & \begin{tabular}[c]{@{}l@{}}\texttt{google.co.th} (Google Thailand), \\ \texttt{thailandpost.co.th} (Thailand Post)\end{tabular} \\ \hline
\textbf{\texttt{.mi.th}} & Military organizations & \begin{tabular}[c]{@{}l@{}}\texttt{rta.mi.th} (Royal Thai Army), \\ \texttt{navy.mi.th} (Royal Thai Navy)\end{tabular} \\ \hline
\textbf{\texttt{.or.th}} & Non-profit organizations & \begin{tabular}[c]{@{}l@{}}\texttt{glo.or.th} (Government Lottery Office), \\ \texttt{set.or.th} (Stock Exchange of Thailand)\end{tabular} \\ \hline
\textbf{\texttt{.in.th}} & Individuals and general organizations & bnn.in.th (BaNANA IT products) \\ \hline
\textbf{\texttt{.net.th}} & Network-related organizations & \begin{tabular}[c]{@{}l@{}}\texttt{uni.net.th} (Office of Information Technology for Educational Development), \\ \texttt{cat.net.th} (National Telecommunications Public Company Limited)\end{tabular} \\ \hline
\end{tabular}
\end{table*}
\section{Methodology}
\label{sec:method}
In this section, we will outline the details and methodology employed in the development of a system designed for the collection of personal data.

\subsection{Personal Data Searching}
The collection of personal data in this system relies on search results retrieved from search engines. Our system supports searches and data aggregation from Google and Bing, two of the most widely used search engines, based on a survey conducted in 2024~\cite{pop_search_engines}. To facilitate this, we have developed a Search Engine Crawler, which is responsible for sending search queries to search engines and collecting the results. The collected data consists of the following components: (1) Search Keywords, (2) URLs resulting from the search, and (3) the timestamp indicating when the search was conducted and the results were retrieved. These data are then stored in our search result database.

Our Search Engine Crawler can be configured with the following parameters: (1) Search Keyword(s), (2) Preferred Search Engine(s) (either Google and/or Bing), (3) Search Delay, which controls the interval between search queries and the collection of results from each search results page, (4) Download Timeout and Download Max Retry, which set the maximum time and number of attempts to download the result URLs, (5) Maximum Search Result Pages, and (6) Desired File Type for the downloaded results.

\subsubsection{Maximum Number of Search Results}
The maximum number of search results refers to the total number of relevant results returned for a given search query, which are ranked according to their relevance, as discussed in Section~\ref{sec:searchengine}. These results are then distributed across multiple pages of search results. Our experiments indicate that most search engines typically display between 6-10 results per page, although this number may vary. The Search Engine Crawler we developed is configurable, allowing users to define the maximum number of pages to crawl, with a default setting of 10 pages. Furthermore, the crawler is capable of detecting the actual number of pages, even if this number is fewer than the user-defined or default settings.

\subsubsection{Search Result Retriever and Collection}
Upon retrieving the search results, our system utilizes the Search Result Downloader to collect the URLs from each result page. The downloader operates according to two key parameters: Download Timeout and Download Max Retry, which govern the maximum duration and the number of retry attempts for downloading the result URLs. Once all results on a given page have been successfully downloaded or the timeout is reached, the downloader enters a brief idle state before proceeding to fetch the next page, as determined by the specified search delay. This delay is designed to simulate real user behavior, as users typically spend some time reviewing the search results before advancing to the next page.

\subsubsection{Search Result Information and File Download}
After each download, the system records key information in the database, including the search keyword, the search result page on which the item appears, the download URL, the download status, and the date and time of retrieval. Additionally, the system calculates a SHA-256 hash for each file to facilitate integrity checks and support subsequent comparisons. This hash is particularly useful for identifying duplicate files that may appear across different search results or websites.

The downloader further verifies that the downloaded files match the expected file types, ensuring the accuracy and completeness of the retrieved content. This verification is accomplished by examining the file extension specified in the Content-Disposition header, a standard HTTP header that indicates the file should be downloaded along with its name and extension. Once verified, the file is renamed with the corresponding hash value of its content, and both the hash and file type are recorded in the database, along with the previously noted information.

\subsection{Search Keyword Selection}
In order to obtain search results that capture the most personal data of Thai citizens, the selection of search keywords is of paramount importance. The following principles guide the selection of search keywords:

\subsubsection{Keywords Related to National ID Numbers}
Examples of such keywords (all translated from Thai) include 
\texttt{ID card number},
\texttt{National ID number} 
\texttt{ID card code}, 
\texttt{number}, 
\texttt{ID card}, etc. 
These search terms are aimed at locating information related to National ID numbers, typically used in conjunction with advanced search techniques such as quotes to ensure the search results must include these specific terms. Additionally, advanced search operators such as \texttt{"OR"} can be used to encompass multiple keywords within a single search query. For example: 
\texttt{"ID card number" OR "National ID number"}.

\subsubsection{File Type-Specific Keywords (filetype)}
Examples of such search terms include \texttt{"filetype:xls"}, \texttt{"filetype:pdf"}, etc. Personal data is often stored in file formats such as PDF, XLS, and DOC files. Specifying the file type in a search query helps to locate information contained within these files, which are commonly uploaded to various websites. Our crawler currently restricts searches to only three file types, aligning with its capability to read and extract data from these formats:
(1) PDF documents (\texttt{.pdf})
(2) Spreadsheet documents (\texttt{.xls}, \texttt{.xlsx})
and (3) Word Processing documents (\texttt{.doc}, \texttt{.docx}).

\subsubsection{Search Keywords Based on National ID Number Prefixes}

Given the specific structure of National ID numbers, especially the first five digits, as detailed in Section~\ref{sec:thai_id_num}, these digits can be used to create targeted search keywords. Examples are provided in Table~\ref{tab:search_keyword_id} to illustrate how searches can identify individuals in specific districts, sub-districts, or provinces. Furthermore, advanced search techniques using Quotes ensure that the results contain the specific number pattern and its sequence.

To comprehensively search for personal data across all provinces, the first digit of the National ID number is used as `1', while digits 2-5 correspond to the district code for every province in Thailand. These structured searches help gather extensive personal data.

\begin{table*}[t]
\centering
\caption{Example of Search Keywords Based on National ID Number Prefixes}
\label{tab:search_keyword_id}
\begin{tabular}{|l|l|}
\hline
\textbf{Search Keyword} & \textbf{Meaning of Search Keywords Using Number Prefixes} \\ \hline
\texttt{"1-1001-"} & \begin{tabular}[c]{@{}l@{}}This search term targets \\ National ID numbers for individuals born on or after January 1, 1984 (digit 1) \emph{AND} \\ National ID numbers of individuals born in the Phra Nakhon District of Bangkok (digits 2-5).\end{tabular} \\ \hline
\texttt{"3-2007-"} & \begin{tabular}[c]{@{}l@{}}This search term targets\\ National ID numbers for individuals born before January 1, 1984 (digit 1) \emph{AND} \\ National ID numbers of individuals born in the Si Racha district of Chonburi province (digits 2-5).\end{tabular} \\ 
\hline
\end{tabular}
\end{table*}

\subsubsection{Keywords Related to Personal Names}
Examples (all translated from Thai) include \texttt{"Mr."}, \texttt{"Mrs."}, \texttt{"Miss"}, \texttt{"Ms."}, \texttt{"Master"/"Mstr"} (boy), \texttt{"Miss"} (girl), etc. These keywords are intended to refine search results to include data related to personal names by specifying common prefixes used in Thai culture.

\subsubsection{Keywords Identifying Website Types}
Examples include \texttt{"site:go.th"} and \texttt{"site:ac.th"}. These keywords are used to restrict search results to specific types of websites, such as government or academic sites. For example, \texttt{"site:go.th"} filters search results to only those originating from government websites, as outlined in Section~\ref{sec:background}.

\subsubsection{Other Related Search Terms}
Examples (all translated from Thai) include ``name list", ``list of senior citizens", ``list of those eligible for allowances", and ``certificate of tax withholding". These search terms were derived from analysis of search results and personal data found, and frequently recurring results were used to create new search keywords to further gather personal data. For example, some keywords target the list of individuals receiving government allowances, and this term was found to appear frequently in search results, proving useful in collecting personal data.
To maximize the potential for gathering personal data, a combination of these search keywords was used. The search results, along with the used search terms, were stored for future reference. Examples include:

\begin{itemize}
    \item \texttt{site:ac.th filetype:xlsx OR \\filetype:xls "number" "citizen" "Mr."}
    \item \texttt{filetype:xls "National ID number" "name"}
    \item \texttt{filetype:pdf "1-3501-" "number" \\"citizen"}
    \item \texttt{"certificate of tax withholding" \\filetype:pdf site:go.th ("Miss" AND "Mr.")}
    \item \texttt{filetype:pdf ("ID card number" OR \\"National ID number" OR "number") "list"
    \\"1-1001-"}
\end{itemize}

\subsection{Personal Data Collection}
The collection of personal data begins with the extraction of text from files, followed by a thorough review of the extracted content to identify relevant personal data for storage. To evaluate the potential risk of personal data leakage, we have specifically focused on extracting National ID numbers as a representative marker for such leakage. The development process for this procedure is outlined as follows:

\subsubsection{Text Extraction}
Text extraction is the process of retrieving or extracting text from files for use or analysis. This is achieved through the creation of a parser tailored to the specific type of file collected. The parsers developed for this purpose are all based on open-source libraries and have been designed to handle the following file types:

\textbf{PDF Documents (.pdf):} The PDF parser was developed using various PDF parser libraries, including (1) PyPDF2~\cite{pypdf2}, (2) pdfplumber~\cite{pdfplumber}, and (3) pytesseract~\cite{pytesseract}. These libraries were chosen for their effectiveness in extracting Thai text. PyPDF2 and pdfplumber employ methods to extract text and characters directly from the file, ensuring high speed. However, if the PDF file is in image format, such as one created by scanning a document, PyPDF2 and pdfplumber will be unable to extract the text, as they do not support image-based text extraction. To address this, pytesseract is also used to extract text. Pytesseract is an OCR (Optical Character Recognition) tool developed from Google’s Tesseract OCR Engine, which uses machine learning techniques for text and word recognition within images. The disadvantage of this method is that it requires more time for extraction compared to the other two methods. Our developed PDF parser extracts text using all three methods to ensure the maximum amount of text is collected.

\textbf{Spreadsheet Documents (.xls, .xlsx):} The Spreadsheet (or Excel) parser was developed using the Pandas library and its \texttt{read\_excel} function~\cite{pandas}, which allows users to select an engine for reading spreadsheet files. To ensure the Spreadsheet parser can read both old (\texttt{.xls}) and new (\texttt{.xlsx}) formats, the system uses three different engines: openpyxl, odf, and pyxlsb. This combination allows for the best text extraction from spreadsheet files.

\textbf{Word Processing Documents (.doc, .docx):} The Word Processing (or Doc) parser was developed using Document Parser libraries, including (1) docx2txt~\cite{python_docx2txt}, (2) textract~\cite{textract}, and (3) antiword~\cite{antiword}. To ensure the Doc parser can handle both old (\texttt{.doc}) and new (\texttt{.docx}) file formats, our system also uses all the aforementioned libraries to extract the maximum amount of text from the documents.

\subsubsection{National ID Number Validation}
To ensure that only valid National ID numbers are stored, our system performs a validation check on every number extracted from the files during the text extraction process. This check is divided into the following steps:

\textbf{Step 1: Format Search and Verification.}
Since the Thai National ID numbers consist of 13 digits, with a clearly defined format (as discussed in Section~\ref{sec:thai_id_num}). Therefore, we can prepare Regular Expression (RegEx) patterns for searching and extracting numbers that match the expected formats (\eg number of expected digits, position of spaces and hyphens between digits).

To ensure that our system can handle all Thai numerals, we extract these specific numerals and convert them into Arabic numerals (0-9). This conversion guarantees that they can be checked against the defined patterns. Only numbers that match the specified regex patterns are processed in the subsequent steps.

\textbf{Step 2: Checksum Validation.}
Once a number matches the format from Step 1, the system performs a checksum validation using the checksum rule for Thai National ID numbers, as explained in Section~\ref{sec:thai_id_valid}. Numbers that fail the checksum validation are discarded and not stored.

\textbf{Step 3: Prefix Validation.}
This final validation step verifies the first five digits of the number that passed the Step 1 and 2, as discussed in Sections~\ref{sec:thai_id_num} and~\ref{sec:thai_id_valid}, to verify the following:
\begin{itemize}
    \item Digit 1: Must be a number between 1 and 8.
    \item Digit 2-5: Must match the district or sub-district code in Thailand.

\end{itemize}
This validation guarantees that all collected numbers accurately represent a Thai National ID number.

\subsubsection{Personal Data Storage}
After the National ID number has been validated, the valid number is stored in the prepared database. Each stored National ID number is linked to its source, including the search engine used, the search keywords, the date and time of discovery, the URL of the source, and the location of the downloaded document. This information allows for precise tracking of the source of each ID number, aiding in analysis and interpretation of the results. Additionally, this metadata may be useful in case the information needs to be disclosed for legal requests or other regulatory purposes.

\textbf{Data Privacy.} We are committed to protecting the privacy of all individuals whose data was collected during this research. The dataset is securely stored in an offline environment and safeguarded using industry-standard authentication protocols. Access is strictly limited to authorized personnel only. Under no circumstances will any personally identifiable information be disclosed. Further details regarding our ethical practices and responsibilities are provided in Section~\ref{sec:ethics}.
\section{Adversary Model and Economic Analysis}
\label{sec:adversary_model}
This section establishes the adversary model used to assess the threat landscape. We categorize potential attackers by their technical capabilities and the financial costs associated with large-scale data acquisition.

\subsection{Adversary Capabilities}
We categorize potential attackers based on their technical resources and the \textit{realistic harm} they can inflict using the exposed data.

\textbf{Opportunistic Adversaries:} Individual actors or ``script kiddies'' with limited technical resources. Their capability is restricted to manual or semi-automated search engine queries (Dorking). While the barrier to entry is near-zero, the harm is typically limited to small-scale identity theft or individual harassment.

\textbf{Systematic Scrapers:} Organized groups with the infrastructure to perform bulk scraping and data indexing. These actors can automate the extraction of large numbers of identifiers and cross-reference them with other leaked databases (\eg, historical bank breaches). This enables \textit{systemic harm}, such as building comprehensive ``Personas'' for large-scale financial fraud or state-level profiling.

\subsection{Economic Analysis}
The effort required for exploitation is categorized by the ``cost of acquisition,'' which in this context is effectively negligible. This is primarily due to the ease of retrieving indexed search results.

\textbf{Web Scraping Retrieval:} An adversary can implement a search engine crawler using headless automated web engines such as Selenium\footnote{Selenium: \url{https://www.selenium.dev}}, allowing results to be retrieved gradually to bypass bot-detection mechanisms. However, as search engines increasingly implement stricter protections to prevent unauthorized scraping for AI training~\cite{searchguard}, this method may face scalability challenges in the long term.

\textbf{API-Based Retrieval:} Alternatively, official or third-party Search Engine Results Page (SERP) APIs significantly lower the technical barrier. As of 2026, the Google Custom Search API~\cite{goog_search_custom} provides 100 free daily queries, with additional requests costing \$5 per 1,000 queries. Similarly, third-party services such as SerpAPI\footnote{SerpAPI: \url{https://serpapi.com}} offer tiered pricing ranging from \$25 to \$275 per month for 1,000 to 30,000 searches, respectively. These price points demonstrate that an attacker could aggregate over a million sensitive records for a financial investment of less than a few hundred dollars.

\section{Analysis of Personal Data Leak}
\label{sec:eval}


This section provides a systematic analysis of the collected personal data and evaluates its impact on the security landscape of Thai citizens. Our methodology—conducted over a three-month window from March to May 2024, utilized 619 targeted search queries to identify 6,097 unique URLs. From these, we successfully extracted \note{1,263,268} unique National ID numbers from 6,004 documents. 

To put this in perspective, this figure represents approximately 2\% of Thailand’s total population (65.9 million) \cite{th_population}. This volume of exposure suggests a significant systemic failure in data protection, as these identifiers are the primary keys for accessing government and financial services in Thailand. The following analysis categorizes these exposures by (1) Download File Type, (2) Registered Domain, and (3) Top-Level Domain and Ownership to identify the structural drivers of this leakage. 

\subsection{Threats to Validity}
\label{subsec:validity}
Before detailing the results, it is necessary to address the inherent limitations and potential biases of the data collection process:

\textbf{Search Engine and Algorithmic Bias:} The reliance on Google and Bing introduces platform-specific biases. Search results are governed by proprietary ranking algorithms and search filters, which may exclude certain exposed records from our view.

\textbf{Temporal Snapshot:} Data collection occurred over a three-month snapshot. Because search engine indices are dynamic, our results reflect a specific temporal window and do not account for data that may have been de-indexed or exposed outside this period.

\textbf{Lower Bound Estimation:} Most importantly, the figures presented in this section should be interpreted as a \textit{lower bound} of the total exposure. Our methodology only captures indexed ``surface web" documents; it does not account for the Deep Web, un-indexed databases, or files protected by \texttt{robots.txt} that remain accessible via direct links.

\subsection{Download File Types}

Analysis of the downloaded file types (Table~\ref{tab:file_type}) reveals a significant correlation between technical format and the volume of exposed PII. While PDF and Word documents are more numerous in terms of unique URLs, they typically contain ``narrative'' or localized records, such as individual forms or small individual lists. In contrast, spreadsheet formats are structurally optimized for bulk data aggregation.

From a security perspective, this disparity represents a critical \emph{Aggregation Risk}. Although spreadsheets account for only 42.5\% of files containing personal data, they are responsible for over 81\% of the total extracted National ID numbers. Such exposure often stems from \emph{Security Misconfigurations (OWASP A02:2025)}~\cite{owasp2025}, specifically where default server settings allow directory listing or fail to implement proper access control headers on administrative staging folders. 

\begin{table}[t]
\caption{Number of Downloaded Files and National ID Numbers Collected from the Internet Categorized by File Type}
\label{tab:file_type}
\footnotesize
\setlength{\tabcolsep}{3pt} 
\centering
\begin{tabular}{|p{1.7cm}|p{0.9cm}|l|p{2cm}|p{1.7cm}|}
\hline
\textbf{File Type} & \textbf{File Ext.} & \textbf{URLs} & \makecell[l]{\textbf{Downloaded} \\ \textbf{Files} \\ \textbf{Containing} \\ \textbf{Personal Data}} & \makecell[l]{\textbf{Unique}\\ \textbf{National IDs}}\\ \hline
Spreadsheet & .xls, .xlsx & 2,370 & 1,467 & 1,032,236 \\ \hline
PDF & .pdf & 3,298 & 1,938 & 241,093 \\ \hline
Word & .doc, .docx & 429 & 48 & 4,062 \\ \hline
\textbf{Total} &  & 6,379 & 3,453 & 1,263,268 \\ \hline
\end{tabular}
\smallskip \\ 
\raggedright \footnotesize \textit{``Unique National IDs'' refers to unique Thai National ID numbers extracted from the datasets.}
\end{table}

\begin{cbox}{Finding 1: A significant number of Thai individuals have had their National ID numbers and personal data exposed online.}

\begin{itemize}
    \item \textbf{Systemic Scale:} Over three-months period (March to May 2024), our system gathered personal data through search queries, resulting in the extraction of 1.2 million National ID numbers accounting for nearly 2\% of the Thai population. 
    \item \textbf{Aggregation Risk:} While spreadsheet documents represent roughly 42\% of the downloaded files containing personal data, they account for over 81\% (\note{$1,032,236$}) of the total extracted National ID numbers. This disparity highlights that the exposure of tabular, bulk-aggregated data is the primary driver of large-scale PII leakage.
\end{itemize}
\end{cbox}

\subsection{Registered Domain}

This section examines the registered domains serving as the primary vectors for PII exposure, revealing a significant \emph{institutional concentration of risk}. As shown in Table~\ref{tab:top10_domains}, the top 10 domains alone account for over 547k exposed National ID numbers; however, a critical interpretive finding is the \emph{Cross-Domain Correlation} between high-sensitivity data and non-government TLDs. Specifically, the most severe leak (Rank 1, 112k National ID numbers) originated from a \texttt{.com} domain (\textit{pokkrongnakhon.com}), while the fourth largest (60k National ID numbers) occurred on a \texttt{.org} domain (\textit{chpao.org}), both of which are owned by government agencies. This indicates that sensitive civil registries are frequently migrated to commercial or third-party hosting environments that may lack the rigorous security oversight of the official \texttt{.go.th} infrastructure. Consequently, the following section provides a deeper analysis of how these institutional failures manifest across different top-Level domains to identify broader patterns of exposure.

\begin{table*}[ht]
\caption{Top 10 Registered Domain with the Highest Number of National ID Numbers Collected}
\label{tab:top10_domains}
\centering
\footnotesize
\setlength{\tabcolsep}{3pt} 
\begin{tabular}{|c|l|p{7.8cm}|l|p{1.8cm}|l|p{1.3cm}|}
\hline
\textbf{Rank} & \textbf{Domain Name} & \textbf{Registered Domain Owner} & \textbf{URLs} & \textbf{Downloaded Files} & \textbf{FQDNs} & \textbf{Unique National IDs} \\ \hline
1 & pokkrongnakhon.com & Nakhon Si Thammarat Provincial Administration & 12 & 12 & 1 & 112,048 \\ \hline
2 & nfe.go.th & Department of Learning Encouragement & 49 & 49 & 15 & 92,433 \\ \hline
3 & cdd.go.th & Department of Community Development & 81 & 81 & 21 & 70,075 \\ \hline
4 & chpao.org & Chaiyaphum Provincial Administration Organization & 1 & 1 & 1 & 60,626 \\ \hline
5 & mkarea3.go.th & Maha Sarakham Primary Educational Service Area Office (Area 3) & 1 & 1 & 1 & 47,187 \\ \hline
6 & fisheries.go.th & Department of Fisheries & 74 & 74 & 2 & 44,208 \\ \hline
7 & rta.mi.th & Royal Thai Army & 9 & 9 & 4 & 32,783 \\ \hline
8 & ccs2.go.th & Chachoengsao Primary Educational Service Area Office (Area 2) & 3 & 3 & 1 & 29,754 \\ \hline
9 & navy.mi.th & Royal Thai Navy & 10 & 10 & 5 & 29,062 \\ \hline
10 & edudev.in.th & Educational Data System Development Center & 2 & 2 & 1 & 29,039 \\ \hline
\end{tabular}
\end{table*}

\subsection{Top-Level Domain Distribution and Ownership}
\label{sec:tld}

Table~\ref{tab:top_tlds} categorizes the exposure by TLD, highlighting significant variations in how sensitive data is distributed across the Thai web ecosystem. The \texttt{.go.th} domain, the official TLD for Thai government agencies, hosts the highest density of exposure, with over 756k unique National IDs. However, a critical interpretive finding is the substantial volume of IDs discovered on commercial (\texttt{.com}), academic (\texttt{.ac.th}), and organization (\texttt{.org}) TLDs, which collectively expose over 380k unique National IDs.

Analysis of these results reveals a structural correlation between decentralized administration and TLD selection. While central departments generally adhere to \texttt{.go.th} protocols, many local government entities (provincial, district, and subdistrict administrations) frequently register domains under commercial TLDs like \texttt{.com} or \texttt{.in.th} for perceived ease of deployment. This creates a fragmented security perimeter where official civil registries, often containing the same depth of PII as centralized databases, are hosted on third-party infrastructure. This suggests that the vulnerability is not a flaw in the \texttt{.go.th} infrastructure itself, but rather a systemic lack of centralized security oversight for decentralized administrative units. In the following sections, we provide a granular analysis of the specific TLDs and organizational sectors that serve as primary vectors for PII exposure. By scrutinizing these recurring patterns of leakage, we aim to identify structural vulnerabilities and provide actionable recommendations to mitigate future data exposure.

\begin{table*}[ht]
\caption{Number of Download URLs, Files, and National ID Numbers Collected Categorized by TLD}
\label{tab:top_tlds}
\centering
\footnotesize
\setlength{\tabcolsep}{3pt} 
\begin{tabular}{|l|l|l|l|l|l|}
\hline
\textbf{TLD} & \textbf{Downloaded Files} & \textbf{FQDNs} & \textbf{Registered Domains} & \textbf{Unique National IDs} & \textbf{\begin{tabular}[c]{@{}l@{}}Registered Domain Examples\end{tabular}} \\ \hline
go.th & 2,305 & 983 & 776 & 756,278 & nfe.go.th, cdd.go.th,mkarea3.go.th \\ \hline
com & 159 & 76 & 66 & 167,236 & pokkrongnakhon.com, tecs4.com, sktcoop.com \\ \hline
ac.th & 579 & 311 & 160 & 155,139 & tupr.ac.th, ubu.ac.th, veis1.ac.th \\ \hline
org & 36 & 18 & 16 & 66,796 & chpao.org, saraburi2.org, cupsakol.org \\ \hline
mi.th & 20 & 10 & 3 & 61,842 & rta.mi.th, navy.mi.th, tdc.mi.th \\ \hline
in.th & 37 & 18 & 18 & 48,586 & edudev.in.th, ssk.in.th, abt.in.th \\ \hline
N/A* & 33 & 1 & 19 & 15,411 & 122.154.253.83, 122.155.168.174, 203.157.184.6\\ \hline
or.th & 201 & 38 & 34 & 6,057 & baac.or.th, nfcrbr.or.th, mea.or.th \\ \hline
net & 14 & 11 & 10 & 5,763 & utdone.net, kkict.net, phsc.net \\ \hline
ac & 9 & 7 & 1 & 1,706 & thai.ac \\ \hline
co.th & 53 & 19 & 17 & 510 & pea.co.th, skybook.co.th, pwa.co.th \\ \hline
Others & 7 & 5 & 5 & 73 & 1stdirectory.co.uk, thaiconsulate.jp, \\ \hline
\end{tabular}
\end{table*}


\begin{cbox}{Finding 2: Majority of exposed Thai National ID Numbers were collected from government websites.}
\begin{itemize}
    \item \textbf{PII Density:} The \texttt{.go.th} domain exhibits the highest data density, accounting for 58.8\% of all exposed IDs. This concentration reflects the centralized nature of government databases, where institutional leaks have a disproportionately higher impact than those in other sectors.
    \item \textbf{Cross-Domain Persistence:} A significant volume of leaks on non-government TLDs (\texttt{.com}, \texttt{.org}) are traced back to local government entities. This suggests that the vulnerability is rooted in public-sector data handling practices rather than the security of the TLD infrastructure itself.
\end{itemize}
\end{cbox}

\subsubsection{TLD: go.th}
\label{sec:tld_go_th}

The \texttt{.go.th} domain represents the core of the Thai government’s digital presence. This section scrutinizes the distribution of PII exposure across various ministerial bodies to identify which institutional sectors serve as the primary sources of leakage. As shown in Table~\ref{tab:go_th}, we identified a significant concentration of risk within three specific ministries: Interior, Education, and Agriculture.

The Ministry of Interior accounts for the largest share of exposure, with over 409k unique National ID numbers. This is interpreted as a result of the ministry's role as the primary custodian of civil and local administrative data. The Ministry of Education follows with approximately 234k unique National ID numbers, while the Ministry of Agriculture and Cooperatives accounts for 86k. 

Collectively, these three ministries are responsible for over 729k unique National IDs, or roughly 96\% of all leaks within the \texttt{.go.th} domain. This high concentration suggests that PII exposure is not a uniform problem across the government, but is specifically clustered in agencies that manage nationwide beneficiary programs, educational registries, and local community datasets. The following subsections provide a granular analysis of these top-tier ministerial exposures to identify the specific document types and administrative processes driving these leaks.

\begin{table*}[]
\caption{Top 10 Government Ministries Registered under TLD ``go.th" with the Highest Number of Exposed National ID Numbers}
\label{tab:go_th}
\footnotesize
\setlength{\tabcolsep}{2pt} 
\begin{tabular}{|c|p{5.2cm}|l|p{9.2cm}|}
\hline
\multicolumn{1}{|l|}{\textbf{Rank}} & \textbf{Domain Owner Ministry} & \textbf{\begin{tabular}[c]{@{}l@{}}Unique \\ National IDs \end{tabular}} & \textbf{Domain Owner Examples} \\ \hline
1 & Ministry of Interior & 409,012 & Community Development Department, Department of Local Administration, Juab Subdistrict Administrative Organization \\ \hline
2 & Ministry of Education & 234,400 & Department of Learning Encouragement, Maha Sarakham Primary Educational  Service Area Office (Area 3), Chachoengsao Primary Educational Service Area Office (Area 2) \\ \hline
3 & Ministry of Agriculture and Cooperatives & 86,218 & \makecell[l]{Department of Fisheries, Department of Agricultural Extension, \\Department of Royal Irrigation} \\ \hline
4 & Ministry of Public Health & 10,251 & Ministry of Public Health, Kantharawichai Hospital, National Institute for Emergency Medicine \\ \hline
5 & Ministry of Transport & 5,553 & Department of Highways, Department of Rural Roads \\ \hline
6 & Ministry of Labour & 5,533 & Ministry of Labour, Department of Employment, Department of Skill Development \\ \hline
7 & Government Agencies not under the Prime Minister's Office, Ministries, or Departments & 3406 & National Office of Buddhism, Royal Thai Police, Anti-Money Laundering Office \\ \hline
8 & Office of the Prime Minister & 1,617 & Office of the Official Information Commission, Open Government Data of Thailand, Office of the Permanent Secretary \\ \hline
9 & Ministry of Finance & 1,279 & \makecell[l]{The Excise Department, The Revenue Department, \\The Customs Department} \\ \hline
10 & \makecell[l]{Ministry of Natural Resources \\and Environment} & 1,198 & Ministry of Natural Resources and Environment, Department of Groundwater Resources, Department of Mineral Resources \\ \hline
\end{tabular}
\end{table*}

\begin{figure}[t]
    \centering
    \includegraphics[width=1\columnwidth]{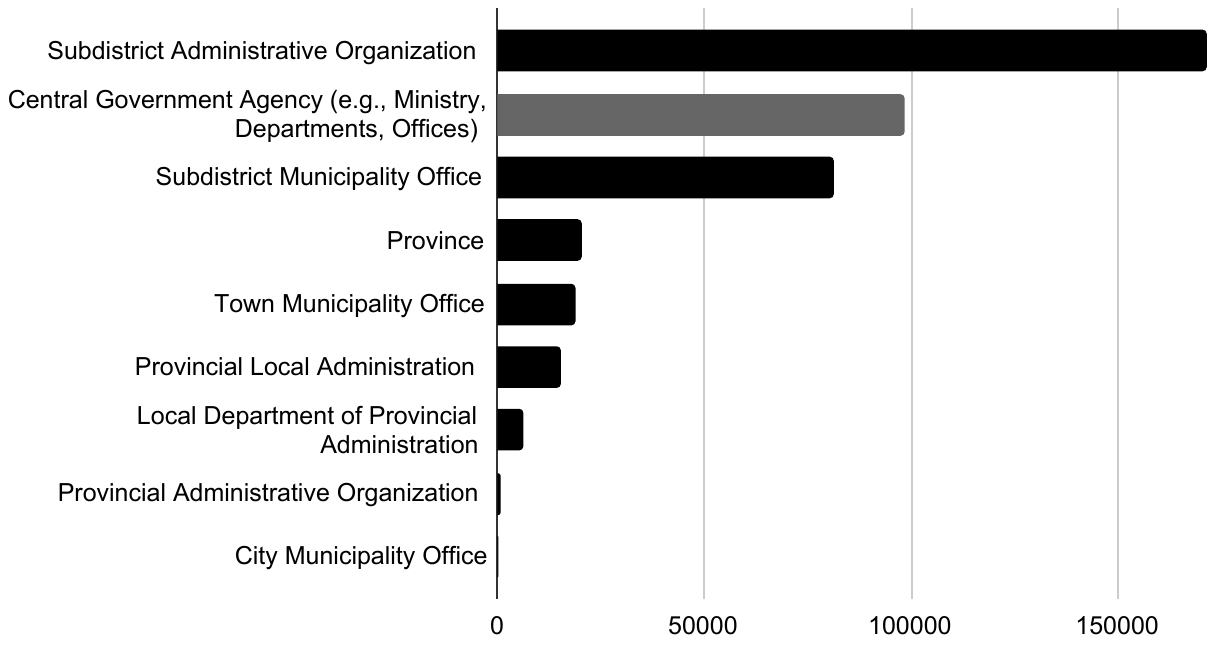}
    \caption{Distribution of Exposed National ID Numbers by Ministry Division within the Ministry of Interior}
    \label{fig:mnt_interior}
\end{figure}

\textbf{Ministry of Interior.} Figure~\ref{fig:mnt_interior} presents the number of National ID numbers associated with government agency domains under various divisions of the Ministry of Interior. Notably, the majority of these domains belong to local government offices, including the Subdistrict Administrative Organization, Subdistrict Municipality Office, Province, Town Municipality Office, Provincial Local Administration Office, Provincial Administrative Organization, and City Municipality Office, collectively accounting for over 76\% (314k) of all exposed National ID numbers within the Ministry. However, central government agencies, \eg~central government offices, also contribute significantly to the total number of exposed ID numbers, over 98k.

\begin{figure*}[t]
    \centering
    \includegraphics[width=0.7\textwidth]{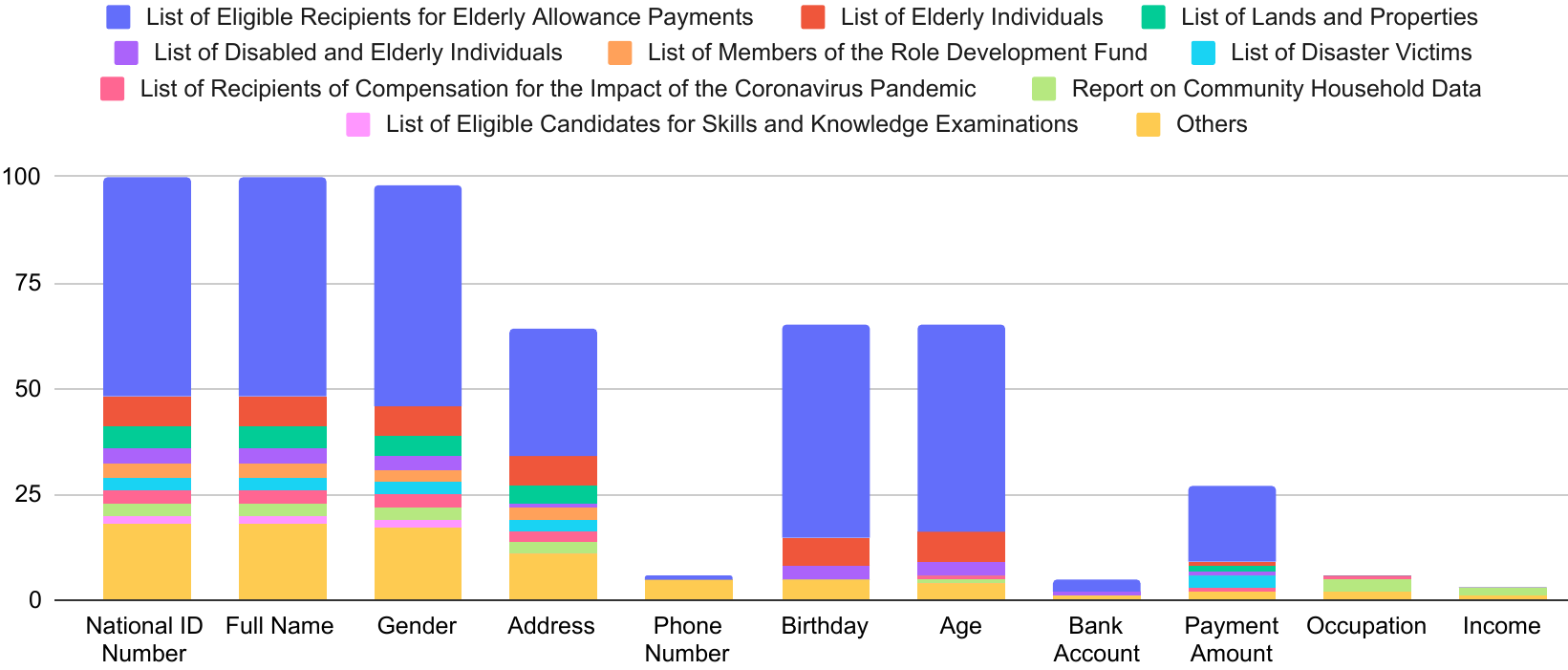}
    \caption{Top 10 Most Frequently Occurring Document Title in the 100 Files Containing the Highest Number of Exposed National ID Numbers from Domains under the Ministry of Interior}
    \label{fig:mnt_interior2}
\end{figure*}

To gain a deeper understanding of the root causes of these leaks, a manual audit of the top 100 high-impact files was conducted (Figure~\ref{fig:mnt_interior2}). Our analysis reveals a recurring \emph{Institutionalized Disclosure Pattern}: the majority of leaks are not the result of malicious breaches, but rather the publication of official administrative registries. Notably, over 60\% of these files comprise lists of vulnerable populations including elderly allowance recipients, persons with disabilities, and disaster victims.

From a security perspective, these documents facilitate \emph{High-Fidelity Identity Reconstruction}. By publishing full names, National IDs, bank accounts, and home addresses alongside sensitive status indicators (\eg disability or pregnancy), these agencies provide malicious actors with a complete profile for targeted social engineering. This is particularly concerning given the growing cybercrime landscape in Thailand; the exposure of phone numbers (found in 5\% of the sampled documents) and income levels directly enables the ``call center" scamming syndicates currently targeting Thai citizens. 

Ultimately, these findings suggest a fundamental conflict in local administrative processes: the pursuit of public transparency regarding government spending and subsidies currently lacks the technical safeguards (such as data masking or access-controlled portals) necessary to protect the constitutional privacy of the beneficiaries.

\textbf{Ministry of Education.}
The Ministry of Education ranks as the second-largest source of PII exposure, characterized by a distinct demographic risk, the large-scale leakage of data belonging to minors. As detailed in Table~\ref{tab:mnt_education}, over 133k unique National IDs were exposed through the Office of the Basic Education Commission, primarily via Primary and Secondary Educational Service Area Offices. For example, sensitive student registries from Prathom (grades 1–6) and Mathayom (grades 7–12) levels were found publicly indexed in regions such as Sukhothai and Maha Sarakham. This is particularly concerning as these data involve not only personal information but also the personal details of minors, which means the leakage affects children from a young age.

Mirroring the patterns observed in the Ministry of Interior, the leakage is driven by decentralized administrative systems. Local educational commissions and vocational offices regularly upload student lists such as enrollments and registries, without adequate data-masking protocols. A particularly egregious discovery was the exposure of data from Special Education Centers, which links National IDs to sensitive indicators of special needs or disabilities.



\begin{table*}[]
\caption{Ministry Divisions under the Ministry of Education with the Highest Number of Exposed National ID Numbers}
\label{tab:mnt_education}
\footnotesize
\setlength{\tabcolsep}{2pt} 
\begin{tabular}{|l|l|p{10cm}|}
\hline
\textbf{Ministry Division} & \textbf{\begin{tabular}[c]{@{}l@{}}Unique \\ National IDs \end{tabular}} & \textbf{Domain Owner Examples} \\ \hline
Office of the Basic Education Commission & 133,396 & Sukhothai Secondary Educational Service Area Office, Chiang Mai Primary Educational Service Area Office (Area 4), Maha Sarakham Secondary Educational Service Area Office \\ \hline
\begin{tabular}[c]{@{}l@{}}Central Government Agencies \end{tabular} & 94,540 & Department of Learning Encouragement \\ \hline
\begin{tabular}[c]{@{}l@{}}Office of the Private \\ Education Commission\end{tabular} & 3,677 & Narathiwat Office of the Private Education, Songkhla Office of the Private Education, Yala Office of the Private Education \\ \hline
\begin{tabular}[c]{@{}l@{}}Office of the Vocational \\ Education Commission\end{tabular} & 2,284 & Office of the Vocational Education Commission \\ \hline
Provincial Education Office & 652 & Rayong Provincial Educational Office, Trang Provincial Educational Office, Loei Provincial Educational Office \\ \hline
Special Education Center & 53 & Chonburi Special Education Center (Area 12) \\ \hline
District Learning Encouragement Center & 1 & Sung Men District Learning Encouragement Center \\ \hline
\end{tabular}
\end{table*}

\textbf{Ministry of Agriculture and Cooperatives.}
The Ministry of Agriculture and Cooperatives represents the third-largest institutional source of PII exposure, characterized by the leakage of comprehensive agrarian registries. As shown in Table~\ref{tab:mnt_agriculture}, the Department of Fisheries and the Department of Agricultural Extension are the primary contributors, collectively exposing over 62k unique National IDs. 

Our qualitative analysis reveals a unique and highly invasive data pattern, the coupling of PII with geospatial and asset Metadata. For example, registries published by the Department of Fisheries do not merely list National ID numbers and names; they include precise geographical coordinates (latitude and longitude), farm dimensions, and livestock types. From a security perspective, this creates a \emph{Physical-Digital Risk Linkage}, where a malicious actor can not only identify an individual but also remotely assess their physical assets, land value, and precise location.

Furthermore, the discovery of agricultural subsidy lists containing bank account information and the exposure of retired officer registries mirrors the patterns in the Ministry of Interior. The inclusion of professional metadata, such as occupation and farm registration dates, enables sophisticated spear-phishing campaigns, where scammers can pose as government officials to ``verify" agricultural grants or land reform status, leveraging the victim's data to build trust.


\begin{table}[t]
\caption{Top 10 Ministry Divisions under the Ministry of Agriculture and Cooperatives with the Highest Number of Exposed National ID Numbers}
\label{tab:mnt_agriculture}
\footnotesize
\setlength{\tabcolsep}{2pt} 
\begin{tabular}{|c|p{5.7cm}|l|}
\hline
\textbf{Rank} & \textbf{Ministry Division} & \textbf{\begin{tabular}[c]{@{}l@{}}Unique \\ National IDs\end{tabular}} \\ \hline
1 & Department of Fisheries & 44,008 \\ \hline
2 & Department of Agriculture Extension & 18,907 \\ \hline
3 & Department of Royal Irrigation & 8,154 \\ \hline
4 & Office of the Permanent Secretary for Ministry of Agriculture and Cooperatives & 4,448 \\ \hline
5 & Department of Livestock Development & 3,192 \\ \hline
6 & Department of Cooperative Auditing & 2,292 \\ \hline
7 & Department of Sericulture & 1,770 \\ \hline
8 & Department of Land Development & 1,521 \\ \hline
9 & Office of Agricultural Land Reform & 1,143 \\ \hline
10 & Department of Royal Forest & 955 \\ \hline
\end{tabular}
\end{table}



\textbf{Ministry of Public Health.}
The Ministry of Public Health exhibits a lower total volume of unique National ID numbers (approx. 10.5k) compared to other ministries, yet the qualitative sensitivity of the exposed data is significantly higher. As shown in Table~\ref{tab:mnt_health}, leaks are distributed across central agencies and provincial hospitals, representing a direct compromise of the intersection between civil identity and healthcare metadata.

From the top 10 files that exposed the highest number of National ID numbers, we discovered several alarming instances of personal data leak. For instance, we identified a list of over 2,200 government officers, which included their full names, National ID numbers, workplace and salaries. Additionally, we found approximately 3,000 Thai withholding tax documents, which typically contained full names, National ID numbers, addresses, and wages. The most critical finding, however, is the exposure of pediatric registries containing the National ID numbers of approximately 500 children. Unlike previous educational leaks, these records include \emph{clinical metadata}, such as medication information. The exposure of a minor's medical history alongside their permanent National ID number creates a permanent, non-remediable privacy violation. Such data is highly sought after for insurance fraud or sophisticated social engineering.



\begin{table}[t]
\caption{Ministry Divisions under the Ministry of Public Health with Highest Number of Exposed National ID Numbers}
\label{tab:mnt_health}
\centering
\footnotesize
\setlength{\tabcolsep}{2pt} 
\begin{tabular}{|p{6cm}|l|}
\hline
\textbf{Ministry Division} & \textbf{\begin{tabular}[c]{@{}l@{}}Unique \\National IDs\end{tabular}} \\ \hline
Central Government Agencies & 5,689 \\ \hline
Hospital & 2,635 \\ \hline
Public Organization & 973 \\ \hline
Regional Health Promotion Center & 562 \\ \hline
Local Public Health Office & 530 \\ \hline
Regional Health Provider Office & 113 \\ \hline
\end{tabular}
\end{table}

\textbf{Other Government Ministries}
As presented in the Table~\ref{tab:go_th}, other ministries, including the Ministry of Labour, Ministry of Transport, and various government agencies not directly under the Prime Minister's Office, were found to have exposed approximately 20k National ID numbers. This constitutes about 5\% of the total National ID numbers disclosed across government websites with the .go.th domain. To further analyze the scope of this data breach, we identified the top 10 files containing the highest number of exposed National ID numbers. These files included sensitive information such as lists of retired government officials, individuals receiving government subsidies due to unemployment, monks and their affiliated temples, recent graduates, individuals participating in government project bidding, and individuals who had received government compensation for injuries sustained during political violence. %
Other personal data were also exposed. Some of these documents contained additional sensitive details, including phone numbers, home addresses, height, weight, health conditions, salaries, and ages. Particularly concerning were those documents that linked individuals' personal information to their political affiliations, raising significant privacy and security concerns.

\begin{cbox}{Finding 3: The exposure of personal data on local government websites presents serious privacy and security concerns.}
\begin{itemize}
    \item \textbf{Decentralized Risk in Local Government:} The majority of exposed data originates from the Ministry of Interior, Education, Agriculture, Public Health, and Labour. Notably, the leakage is driven by \emph{local government agencies} (\eg subdistrict offices and provincial service areas), suggesting that decentralized administrative units lack the robust security oversight present in central ministries.
    \item \textbf{Multidimensional Identity Exposure:} In addition to National ID numbers, the datasets include phone numbers, bank details, and income levels. This ``data clustering" creates a high-fidelity profile for each victim, significantly lowering the barrier for highly targeted financial fraud and social engineering (e.g., call center scams).
    \item \textbf{Sensitive Social Indicators:} The exposure of data related to disabilities, pregnancy, and political affiliations represents a profound breach of privacy. Such sensitive indicators could lead to \emph{social discrimination or predatory targeting}, moving the risk from mere identity theft to potential long-term human rights implications.
\end{itemize}
\end{cbox}

\subsubsection{TLD: com}
\label{sec:tld_com}
The \texttt{.com} TLD represents the second-largest vector of exposure, accounting for 13\% (167k) of all unique National IDs. While traditionally reserved for commercial enterprise, a substantial portion of personal data exposure, as highlighted in Finding 1, originates from government agency websites, as presented in Table~\ref{tab:top10_com}. Many of these domains remain active or have not been properly decommissioned after their intended use, for example, Chachoengsao Primary Educational Service Area Office (Area 2) website. As a result, certain files containing sensitive data remain accessible online. Another notable pattern is the storage or hosting of personal information on external websites, such as filethaischool1.com and wordpress.com, or file-sharing services like filesesbuy.com.


\begin{table}[t]
\caption{Top 10 Website Categories of Websites Registered under .com TLD with Highest Number of Exposed National ID Numbers}
\label{tab:top10_com}
\centering
\footnotesize
\setlength{\tabcolsep}{2pt} 
\begin{tabular}{|c|l|l|}
\hline
\textbf{Rank} & \textbf{Website Category} & \textbf{Unique National IDs} \\ \hline
1 & Government and Legal Organizations & 128,955 \\ \hline
2 & Education & 12,979 \\ \hline
3 & Finance and Banking & 8,369 \\ \hline
4 & Dynamic DNS & 6,195 \\ \hline
5 & Business & 6,054 \\ \hline
6 & Health and Wellness & 3,409 \\ \hline
7 & Web Hosting & 695 \\ \hline
8 & Newsgroups and Message Boards & 314 \\ \hline
9 & File Sharing and Storage & 152 \\ \hline
10 & General Organizations & 123 \\ \hline
\end{tabular}
\end{table}

\textbf{Business Sector Leakage and Data Spillover.} 
To understand the extent of personal information exposure within the business sector, we focused our analysis on websites categorized under the Business website category. We identified a concerning trend of government data spillover. Business websites were found hosting documents that should theoretically be restricted to state registries, including property ownership and disability pension recipient lists. In contrast, a more justifiable instance of personal information exposure within the business sector is the publication of employee lists within an organization, as such records are typically maintained for internal administrative purposes.

\subsubsection{TLD: ac.th}
\label{tld_ac_th}

The \texttt{.ac.th} Top-Level Domain, designated for academic institutions, represents 12\% (155k) of the unique National ID numbers in our dataset. To gain a deeper understanding, we categorized the .ac.th domains based on institution types, following the classification by the Ministry of Higher Education, Science, Research, and Innovation (MHESI)~\cite{mhesi_cat}. As shown in Table~\ref{tab:tld_ac_th}, the highest volume of exposure (over 62k IDs) originates from Primary and Secondary Schools (Grades 1–12), followed by autonomous and public universities. This distribution highlights a critical early-stage identity compromise, where sensitive PII is leaked at the very beginning of an individual's civic life.


Furthermore, we identified the top 10 documents with the highest number of exposed National ID numbers and conducted a content analysis to determine additional leaked data. These documents typically contain students' full names, along with their academic level, class, and grade. However, some documents also disclose highly sensitive personal details, including date of birth, nationality, race, religion, disability status, height, weight, blood type, phone number, and home address even for minors under the age of 18. Additionally, some records include parental information, further exacerbating privacy concerns.

Additionally, we observed a concerning pattern where university students who have taken out student loans are identifiable within the leaked data. This is particularly sensitive as it publicly reveals individuals who are in financial debt,
further extending the scope of exposed personal information.

\subsubsection{TLD: org}
\label{sec:tld_org}
The \texttt{.org} TLD is primarily designated for organizations, particularly non-profit entities. Notable websites using the .org extension include the Red Cross and Wikipedia. However, an interesting observation is that a significant portion of this TLD in Thailand are still associated with government sectors. Examples include websites of the Chaiyaphum Provincial Administrative Organization, Saraburi Primary Educational Service Area Office (Area 2), and Local Public Health Offices. Mirroring the vulnerabilities of the \texttt{.go.th} sector, these \texttt{.org} domains frequently host non-anonymized beneficiary registries for allowance payments, confirming a systemic failure in data privacy standards regardless of the TLD.


\textbf{Corporate and Other Organizations.} Beyond government-affiliated .org domains, we also identified personal data exposure in documents hosted by corporate entities and other organizations. These documents contain full names and National ID numbers, along with other sensitive information such as passport numbers and marital status,
further highlighting the widespread risk of personal data leakage across different sectors.

\subsubsection{TLD: mi.th}
\label{sec:tld_mi_th}
The \texttt{.mi.th} TLD is designated for domains associated with the Thai military. Our analysis revealed that .mi.th-registered domains ranked fifth among all identified TLDs in terms of personal data exposure, accounting for 4.8\% of all exposed National ID numbers. The majority of these leaks originated from two primary domain owners: the Royal Thai Army and the Royal Thai Navy, with over 32k and 29k records exposed online, respectively.

Our qualitative analysis of high-impact military files reveals a pattern of institutionalized structural exposure. These files exclusively pertain to military personnel records, containing names, National ID numbers, and military ranks. Notably, some documents also include lists of personnel assigned to specific missions and sub-organizational units, information that could be considered sensitive in terms of military and national security confidentiality.

\subsubsection{Other TLDs and IP Addresses}
\textbf{Other TLDs} such as \texttt{.in.th}, \texttt{.or.th}, and \texttt{.net} account for approximately 62k unique National ID numbers, representing around 5\% of all exposed National ID numbers identified in our analysis. Our findings indicate a recurring pattern observed in other TLDs, where the majority of data exposures originate from the government sector. Specifically, local government agencies that registered their domains outside the \texttt{.go.th} TLD contribute significantly to these leaks. Interestingly while some domains do not officially belong to government agencies such as abt.in.th, thai.ac, and thaischool.in.th, they are typically still connected to the government sector, primarily serving as website hosting services for local government agencies and school websites.

\textbf{IP-address websites} account for 15k unique National ID numbers, representing approximately 1\% of all exposed National ID numbers identified in our analysis. In total, there are 34 IP addresses that contain personal information. We audited the top 10 source IPs with the highest number of exposed National ID numbers and performed a reverse lookup on their IP Whois registration information to gather details about their geolocation and registered owners. Our findings revealed that all 10 IPs are located in Thailand.
In some cases, the domains are linked to government sectors, such as the Ministry of Public Health and the Ministry of Higher Education, Science, Research, and Innovation. %
We also identified that some file also contains full names, dates of birth, ages, and addresses. All these top-10 documents are associated with various government-related duties, such as lists of children in compulsory education and the survey list of individuals with risk behaviors for non-communicable diseases. This further highlights the extensive scope of personal data exposure within government-sector documents.

\begin{cbox}{Finding 4: Most personal data exposures stem from the government sector\, even when hosted on non-government domains.}
\begin{itemize}
    \item \textbf{Public Sector Data Gravity:} While non-government TLDs (\eg~.com, .org, .in.th) exhibit high exposure, the owners are almost exclusively \emph{local government agencies}. This indicates the vulnerability is rooted in public-sector data practices rather than TLD infrastructure.
    \item \textbf{Activity-Driven Exposure:} The highest-density leaks are strongly correlated with mass administrative activities, such as welfare distribution, student registration, and public health tracking. This confirms that bulk-data processing is the primary risk vector for large-scale exposure.
    \item \textbf{Lifecycle Management Failure:} Manual review reveals that many exposed records belong to legacy websites. This indicates a \emph{``long-tail risk''} where data remains indexed long after its operational necessity has passed, highlighting a systemic lack of data decommissioning protocols.
\end{itemize}
\end{cbox}

\begin{table}[t]
\caption{Categorization of ac.th TLD Registered Domains by Institution Type and Their Respective Exposed National ID Numbers}
\label{tab:tld_ac_th}
\centering
\footnotesize
\setlength{\tabcolsep}{2pt} 
\begin{tabular}{|p{6cm}|l|}
\hline
\textbf{Institution Type} & \textbf{\begin{tabular}[c]{@{}l@{}}Unique \\National IDs\end{tabular}} \\ \hline
Primary and Secondary School (Grade 1-12) & 62,929 \\ \hline
Autonomous University & 32,856 \\ \hline
Public University & 18,645 \\ \hline
Rajamangala Universities of Technology & 14,752 \\ \hline
Rajabhat University & 11,591 \\ \hline
Vocational College & 10,172 \\ \hline
\makecell[l]{Non-affliated Academic Institution \\under the MHESI} & 2,774 \\ \hline
Private University & 962 \\ \hline
Private College & 625 \\ \hline
Private Institution & 40 \\ \hline
Open Public Universities & 5 \\ \hline
\end{tabular}
\end{table}

\subsection{Search Keyword and Search Engine}

\begin{figure}[t]
    \centering
    \includegraphics[width=.9\columnwidth]{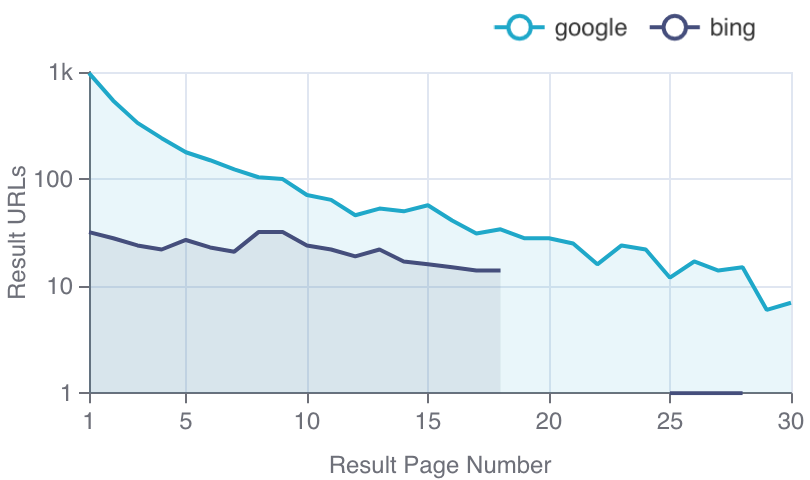}
    \caption{Association between Result URLs containing Personal Information and Result Page Number}
    \label{fig:search_engine}
\end{figure}

In this section, we analyze the correlation between search query structures and the volume of exposed unique National ID numbers. Our results demonstrate that search engines, particularly Google, which accounted for over 99\% of all retrieved data, act as highly efficient, unintentional aggregators of PII. The most effective queries utilized a combination of filetype operators (\eg spreadsheets or PDFs) and Thai-language PII identifiers, suggesting that mass data harvesting requires minimal technical sophistication.



Table~\ref{tab:search_keyword} summarizes the top-performing keyword categories. To ensure ethical compliance and prevent the replication of these findings for malicious use, we have redacted specific numeric patterns and exact keyword strings. The data reveals a synergistic query effect, the combination of contextual Thai-language terms commonly found alongside National ID numbers, domain-specific constraints (\eg~\texttt{site:go.th}) and specific file extensions (\eg~\texttt{.xlsx}) yielded the highest density of unique National ID numbers. 

\begin{table*}[]
\caption{Top 10 Keyword Categories Associated with the Highest Number of Exposed URLs and National ID Numbers}
\label{tab:search_keyword}
\centering
\footnotesize
\setlength{\tabcolsep}{2pt} 
\begin{tabular}{|c|l|l|l|l|}
\hline
\textbf{Rank} & \textbf{Search Keyword} & \textbf{\begin{tabular}[c]{@{}l@{}}Search \\ Engine\end{tabular}} & \textbf{\begin{tabular}[c]{@{}l@{}}URLs\end{tabular}} & \textbf{\begin{tabular}[c]{@{}l@{}}Unique\\ National IDs\end{tabular}} \\ \hline
1 & \tightbox{National ID number term} \tightbox{filetype:xlsx} & google & 40 & 147,345 \\ \hline
2 & \tightbox{filetype:xls OR filetype:xlsx} \tightbox{site:go.th} \tightbox{National ID number term}  \tightbox{Name prefix term} & google & 141 & 115,783 \\ \hline
3 & \tightbox{filetype:xlsx OR filetype:xls} \tightbox{Partial National ID number*} & google & 10 & 107,620 \\ \hline
4 & \tightbox{filetype:pdf} \tightbox{National ID number term}  \tightbox{Partial National ID number*} & google & 150 & 70,085 \\ \hline
5 & \tightbox{filetype:xls OR filetype:xlsx} \tightbox{site:ac.th} \tightbox{National ID number term}  \tightbox{Name prefix term} & google & 80 & 55,886 \\ \hline
6 & \tightbox{National ID number term} \tightbox{filetype:pdf} & google & 104 & 51,672 \\ \hline
7 & \tightbox{National ID number term} \tightbox{filetype:pdf} & google & 128 & 51,267 \\ \hline
8 & \tightbox{filetype:xls OR filetype:xlsx} \tightbox{(National ID number terms)} \tightbox{Partial National ID number*} & google & 1 & 47,187 \\ \hline
9 & \tightbox{site:in.th} \tightbox{filetype:xlsx OR filetype:xls} \tightbox{National ID number term} \tightbox{Name prefix term} & google & 12 & 46,167 \\ \hline
10 & \tightbox{National ID number term} \tightbox{filetype:xls} & google & 34 & 41,913 \\ \hline
\multicolumn{5}{l}{*The exact numbers used have been omitted to prevent the search keywords from being replicated to obtain personal information}
\end{tabular}
\end{table*}


Additionally, we analyzed the ranking distribution of search results to understand where exposed personal data appears on result pages. As illustrated in Figure~\ref{fig:search_engine}, we observed that PII-heavy URLs are not buried in the ``long-tail" of search results; instead, they are frequently indexed within the first 10 pages. This emphasizes that the primary risk is not just the existence of the files, but their high search-rank accessibility, which maximizes exposure to even unsophisticated actors.

\begin{cbox}{Finding 5: Google search is the primary source of exposed National ID numbers\, driven by specific operations and keywords.}
\begin{itemize}
    \item \textbf{High-Value Query Combinations:} Exposure is driven by the synergy between filetype-specific operators and Thai-specific PII keywords. This confirms that targeted Google dorking remains a low-effort, high-reward vector for mass data harvesting.
    \item \textbf{Search Rank Criticality:} The highest concentration of PII appears on the initial search result pages. This indicates that the most significant vulnerabilities are not hidden in the ``long-tail" of search results but are \emph{highly ranked}, maximizing exposure to even unsophisticated actors.
    \item \textbf{Platform Dominance:} While both engines were used, Google's superior indexing of Thai government subdomains makes it the primary discovery source, underscoring the need for search-engine-specific remediation.
\end{itemize}
\end{cbox}

\begin{figure}[t]
    \centering
    \includegraphics[width=.9\columnwidth]{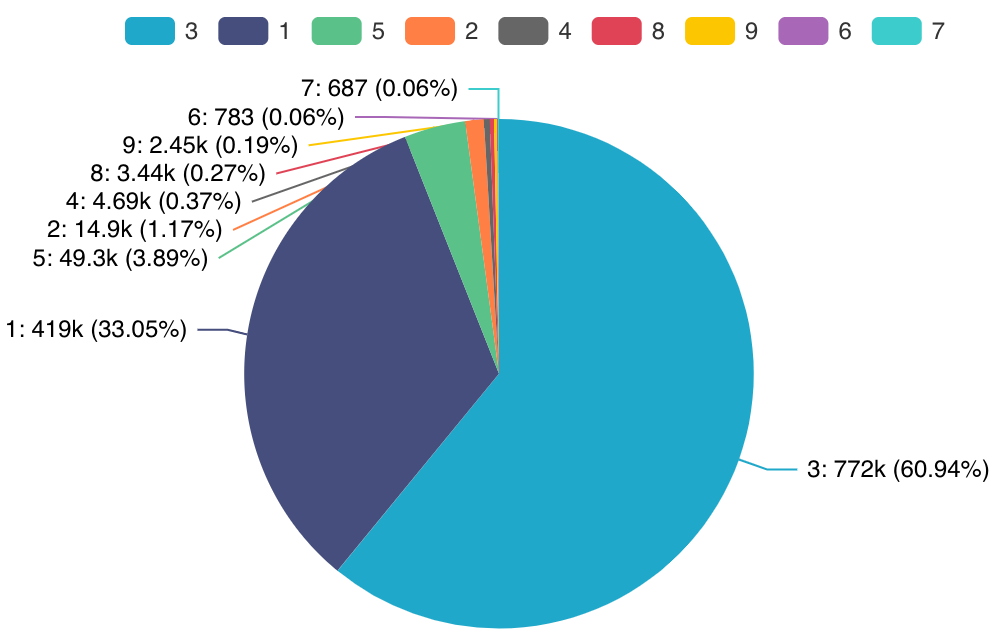}
    \caption{Number of National ID Numbers Grouped by Individual Categories (National ID Number First Digit)}
    \label{fig:cId_type}
\end{figure}

\subsection{Demographic Profile: Individual Category (National ID Number First Digit)}

To move beyond a simple volume count, we analyzed the first digit of the unique National ID numbers to reconstruct the demographic profile of the exposed population. As detailed in Section~\ref{sec:thai_id_num}, the first digit serves as a proxy for age and registration status. As illustrated in Figure~\ref{fig:cId_type}, over 61\% of the collected IDs begin with the digit 3. This cohort represents Thai nationals and long-term residents registered before May 1984, meaning the majority of these individuals are aged 41 or older as of 2025. This age skew is highly significant from a security perspective; this demographic often possesses higher financial assets while simultaneously exhibiting lower digital literacy, making them the primary targets for sophisticated social engineering and ``call center" scams currently prevalent in Thailand. The second-largest cohort (33\%) consists of individuals with the first digit 1, representing Thai nationals born after 1984. 
Combined, these two categories account for over 93\% of the total exposure. This concentration suggests that the current state of PII leakage in Thailand creates a multi-generational security crisis, where the elder population faces immediate financial risk, while the youth population faces a compromised digital future.

\begin{cbox}{Finding 6: Senior individuals and those over 41 represent the majority of exposed personal data.}
\begin{itemize}
    \item \textbf{High-Risk Demographic Skew:} Over 60\% of exposed individuals are aged 41 or older. This demographic skew indicates that the leaks primarily affect a population segment that is often targeted by \emph{social engineering and financial fraud} due to lower digital literacy.
    \item \textbf{Generational Impact:} The inclusion of 33\% of individuals under 41--including student and minor records--highlights a \emph{long-term identity theft risk} that may persist for decades as these individuals enter the workforce and financial systems.
\end{itemize}
\end{cbox}

\subsection{Geographic Distribution: Province and District}

To identify the regions most acutely impacted by PII exposure, we mapped the collected National ID numbers to their geographic origins by extracting digits 2–4, which correspond to the registration province and district. By benchmarking these findings against the 2024 population statistics from the Bureau of Registration Administration~\cite{th_population}, we identified a significant geographic variance in data exposure across Thailand.
As presented in Table~\ref{tab:top10_province1}, Nakhon Si Thammarat recorded the highest absolute volume of exposure (136k unique National ID numbers), while Bangkok followed with 70k. However, when normalized against population size (Table~\ref{tab:top10_province2}), a different risk profile emerges: Satun province exhibited the highest per-capita exposure, with over 14.8\% of its total population affected. This suggests that while major urban centers like Bangkok generate high volumes of data, smaller provincial administrations may suffer from higher institutional vulnerability densities, where a single leak impacts a disproportionately large segment of the local citizenry.

\begin{table}[t]
\caption{Top 10 Provinces with the Highest Number of Personal Data Exposure}
\label{tab:top10_province1}
\centering
\footnotesize
\setlength{\tabcolsep}{2pt} 
\begin{tabular}{|c|l|c|l|l|l|}
\hline
\textbf{Rank} & \textbf{Province} & \textbf{\begin{tabular}[c]{@{}l@{}}Province \\ Code\end{tabular}} & \textbf{\begin{tabular}[c]{@{}l@{}}Unique \\ National IDs\end{tabular}} & \textbf{Population\footnotemark[1]} & \multicolumn{1}{c|}{\textbf{\%}} \\ \hline
1 & \begin{tabular}[c]{@{}l@{}}Nakhon Si\\ Thammarat\end{tabular} & 80 & 136,375 & 1,531,727 & 8.90 \\ \hline
2 & Bangkok & 10 & 70,302 & 5,352,831 & 1.31 \\ \hline
3 & Chaiyaphum & 36 & 65,334 & 1,105,008 & 5.91 \\ \hline
4 & Satun & 91 & 48,057 & 324,390 & 14.81 \\ \hline
5 & Khon Kaen & 40 & 43,602 & 1,768,366 & 2.47 \\ \hline
6 & \begin{tabular}[c]{@{}l@{}}Nakhon\\ Ratchasima\end{tabular} & 30 & 42,135 & 2,615,039 & 1.61 \\ \hline
7 & \begin{tabular}[c]{@{}l@{}}Maha \\ Sarakham\end{tabular} & 44 & 40,776 & 929,056 & 4.39 \\ \hline
8 & Chachoengsao & 24 & 34,699 & 729,218 & 4.76 \\ \hline
9 & Chiang Mai & 50 & 34,445 & 1,635,983 & 2.11 \\ \hline
10 & Kalasin & 46 & 29,749 & 961,369 & 3.09 \\ \hline
\end{tabular}
\end{table}


\begin{table}[t]
\caption[Top 10 Provinces with the Highest Number of Personal Data Exposure]{Top 10 Provinces with the Highest Number of Personal Data Exposure, Ranked by the Percentage of Exposed Data Relative to the Population\footnotemark[1]}
\label{tab:top10_province2}
\centering
\footnotesize
\setlength{\tabcolsep}{2pt} 
\begin{tabular}{|c|l|c|l|l|l|}
\hline
\textbf{Rank} & \textbf{Province} & \textbf{\begin{tabular}[c]{@{}l@{}}Province\\ Code\end{tabular}} & \textbf{\begin{tabular}[c]{@{}l@{}}Unique \\ National IDs\end{tabular}} & \textbf{Population\footnotemark[1]} & \multicolumn{1}{c|}{\textbf{\%}} \\ \hline
1 & Satun & 91 & 48057 & 324390 & 14.81 \\ \hline
2 & \begin{tabular}[c]{@{}l@{}}Nakhon Si\\ Thammarat\end{tabular} & 80 & 136375 & 1531727 & 8.90 \\ \hline
3 & Chaiyaphum & 36 & 65334 & 1105008 & 5.91 \\ \hline
4 & Chachoengsao & 24 & 34699 & 729218 & 4.76 \\ \hline
5 & \begin{tabular}[c]{@{}l@{}}Maha\\ Sarakham\end{tabular} & 44 & 40776 & 929056 & 4.39 \\ \hline
6 & Nan & 55 & 17134 & 468670 & 3.66 \\ \hline
7 & Narathiwat & 96 & 28032 & 822827 & 3.41 \\ \hline
8 & Phatthalung & 93 & 17182 & 519103 & 3.31 \\ \hline
9 & Kalasin & 46 & 29749 & 961369 & 3.09 \\ \hline
10 & Sukhothai & 64 & 17241 & 572575 & 3.01 \\ \hline
\end{tabular}
\end{table}

\textbf{District-Level Saturation.}
To gain deeper insights, we conducted the same statistical analysis at the district level. Table~\ref{tab:top10_district1} presents the top 10 districts with the highest number of personal data exposures identified in our study. Notably, five districts from Nakhon Si Thammarat exhibit exposure rates ranging from 8\% to 19\% of their district populations, while two districts from Satun show the highest exposure rates, ranging from 19\% to 21\%. These findings highlight significant regional disparities in data exposure.

Table~\ref{tab:top10_district2} ranks the top 10 districts based on the percentage of exposed personal data relative to their current population. Notably, Sanam Chai Khet Sub-district Municipality in Chachoengsao province exhibits an exposure rate that exceeds its recorded population. Furthermore, Chiang Yuen Sub-district Municipality and Kosum Phisai Sub-district Municipality in Maha Sarakham province have exposure rates surpassing 50\% of their respective populations, highlighting significant vulnerabilities in these areas. These saturation leaks are interpreted as a failure of local administrative registries, where entire community databases, including historical records and non-resident registrants, have been published en masse. 

\begin{table*}[]
\caption{Top 10 District with the Highest Number of Personal Data Exposure}
\label{tab:top10_district1}
\centering
\footnotesize
\setlength{\tabcolsep}{2pt} 
\begin{tabular}{|c|l|c|l|c|l|l|l|}
\hline
\textbf{Rank} & \textbf{Province} & \textbf{\begin{tabular}[c]{@{}c@{}}Province Code\end{tabular}} & \textbf{District} & \textbf{\begin{tabular}[c]{@{}c@{}}District Code\end{tabular}} & \textbf{\begin{tabular}[c]{@{}l@{}}Unique \\National IDs\end{tabular}} & \textbf{Population\footnotemark[1]} & \multicolumn{1}{c|}{\textbf{\%}} \\ \hline
1 & \begin{tabular}[c]{@{}l@{}}Nakhon Si Thammarat\end{tabular} & 80 & \begin{tabular}[c]{@{}l@{}}Nakhon Si Thammarat City\end{tabular} & 8001 & 17,599 & 161,786 & 10.88 \\ \hline
2 & Satun & 91 & Satun City & 9101 & 14,183 & 66,389 & 21.36 \\ \hline
3 & Nakhon Si Thammarat & 80 & Tha Sala & 8008 & 13,592 & 113,346 & 11.99 \\ \hline
4 & Satun & 91 & La-ngu & 9105 & 13,363 & 68,932 & 19.39 \\ \hline
5 & Chaiyaphum & 36 & Chaiyaphum City & 3601 & 12,246 & 135,262 & 9.05 \\ \hline
6 & Khon Kaen & 40 & Khon Kaen City & 4099 & 12,056 & 98,880 & 12.19 \\ \hline
7 & Maha Sarakham & 44 & Kosum Phisai & 4403 & 10,959 & 107,980 & 10.15 \\ \hline
8 & Nakhon Si Thammarat & 80 & Ron Phibun & 8013 & 10,110 & 61,153 & 16.53 \\ \hline
9 & Nakhon Si Thammarat & 80 & Chawang & 8004 & 9,858 & 52,103 & 18.92 \\ \hline
10 & Nakhon Si Thammarat & 80 & Thung Song & 8009 & 9,593 & 115,660 & 8.29 \\ \hline
\end{tabular}
\end{table*}

\footnotetext[1]{Population Survey as of December 2024~\cite{th_population}.}

\begin{table*}[]
\caption[Top 10 District with the Highest Number of Personal Data Exposure]{Top 10 District with the Highest Number of Personal Data Exposure, Ranked by the Percentage of Exposed Data Relative to the Population\footnotemark[1]}
\label{tab:top10_district2}
\centering
\footnotesize
\setlength{\tabcolsep}{2pt} 
\begin{tabular}{|c|l|c|l|c|l|l|l|}
\hline
\textbf{Rank} & \textbf{Province} & \textbf{Province Code} & \textbf{District} & \textbf{District Code} & \textbf{\begin{tabular}[c]{@{}l@{}}Unique\\National IDs\end{tabular}} & \textbf{Population\footnotemark[1]} & \multicolumn{1}{c|}{\textbf{\%}} \\ \hline
1 & Chachoengsao & 24 & Sanam Chai Khet Sub-district & 2481 & 6983 & 4,231 & 165.04 \\ \hline
2 & Maha Sarakham & 44 & Chiang Yuen Sub-district & 4494 & 3067 & 4,481 & 68.44 \\ \hline
3 & Maha Sarakham & 44 & Kosum Phisai Sub-district & 4496 & 4964 & 8,834 & 56.19 \\ \hline
4 & Chachoengsao & 24 & Thung Sadao Sub-district & 2480 & 1523 & 5,782 & 26.34 \\ \hline
5 & Satun & 91 & Khuan Don & 9102 & 5019 & 22,228 & 22.58 \\ \hline
6 & Surat Thani & 84 & Surat Thani City & 8401 & 4223 & 19,177 & 22.02 \\ \hline
7 & Satun & 91 & Satun City & 9101 & 14183 & 66,441 & 21.35 \\ \hline
8 & Phatthalung & 93 & Khuan Khanun Sub-district & 9395 & 394 & 1,962 & 20.08 \\ \hline
9 & Satun & 91 & La-ngu  & 9105 & 13363 & 69,137 & 19.33 \\ \hline
10 & Nakhon Si Thammarat & 80 & Hua Sai & 8016 & 8178 & 43,277 & 18.90 \\ \hline
\end{tabular}
\end{table*}

\begin{cbox}{Finding 7: Specific provinces and sub-districts exhibit localized ``saturation leaks'' with exposure rates exceeding 50\% of their populations.}
\begin{itemize}
    \item \textbf{Regional Exposure Saturation:} While Nakhon Si Thammarat has the highest volume (136k), Satun exhibits a higher \emph{per-capita risk}, affecting 14\% of its population. This indicates that smaller provinces may face higher systemic exposure relative to their size.
    \item \textbf{Sub-district Vulnerability Peaks:} In specific municipalities (\eg~Sanam Chai Khet, Chiang Yuen), exposure rates \emph{exceed 50\% of the local population}. These ``saturation leaks" suggest that a single administrative error can compromise an entire local community's privacy.
\end{itemize}
\end{cbox}

\subsection{Repeated Exposure: Fragmented Identity Aggregation}
In this section, we analyze the frequency of an individual's personal data appearing across multiple data sources. The aim is to evaluate the extent of repeated exposure, which could heighten 
the risk of further data exposure, enabling potential attackers to gather more detailed information about individuals. By cross-referencing National ID numbers as identifiers across sources, we determine how often an individual's data appears on different sources. Table~\ref{tab:repeated} presents the distribution of National ID numbers exposed across 1 to 15 distinct source URLs, revealing that over 90\% of the exposed data comes from a single source URL. 
%
We analyzed National ID numbers appearing across 15 distinct sources.
We observed that, beyond the basic personal data such as full name, National ID number, and gender, additional data points such as place of work, job position, educational background, salary and contracts were also exposed. This additional information significantly increases the risks associated with data exposure, potentially allowing malicious actors to gain more comprehensive profiles of individuals.

\begin{cbox}{Finding 8: Most personal information exposure comes from a single source\, but repeated leaks across multiple sources heighten the risk of misuse.}
\begin{itemize}
    \item \textbf{Single-Source Criticality:} Over 90\% of the exposed National ID numbers originate from a single source URL, reflecting a significant concentration of data exposure in a limited number of sources.
    \item \textbf{Cross-Source Correlation Risk:} While most individuals appear in only one leak, a subset appears across multiple URLs. This \emph{multi-source exposure} allows attackers to cross-reference fragmented data 
    facilitating more sophisticated social engineering.
\end{itemize}
\end{cbox}

\begin{table}[t]
\caption{Distribution of Exposed National ID Numbers Across Multiple Source URLs}
\label{tab:repeated}
\centering
\footnotesize
\setlength{\tabcolsep}{2pt} 
\begin{tabular}{|l|l|l|}
\hline
\textbf{Source URLs} & \textbf{Unique National IDs} & \multicolumn{1}{c|}{\textbf{\%}} \\ \hline
15 & 5 & 0.0004 \\ \hline
13 & 6 & 0.0005 \\ \hline
12 & 56 & 0.0044 \\ \hline
11 & 7 & 0.0006 \\ \hline
10 & 7 & 0.0006 \\ \hline
9 & 61 & 0.0048 \\ \hline
8 & 79 & 0.0063 \\ \hline
7 & 774 & 0.0613 \\ \hline
6 & 3795 & 0.3003 \\ \hline
5 & 3332 & 0.2637 \\ \hline
4 & 9584 & 0.7584 \\ \hline
3 & 16229 & 1.2843 \\ \hline
2 & 89890 & 7.1138 \\ \hline
1 & 1139443 & 90.2008 \\ \hline
\end{tabular}
\end{table}
\section{Countermeasures and Discussion}
\label{sec:counter}

Our work highlights the privacy implications of personal data exposure in public domains, demonstrating the severity and widespread nature of such incidents.

\subsection{Strengthening Government Data Policies and Governance} 
As demonstrated in Section~\ref{sec:eval}, personal data exposure is both prevalent and large-scale, with the majority of leaks originating from government websites especially those operated by local administrative offices. While publishing certain personal information may support transparency and operational efficiency (\eg~publishing lists of welfare recipients or school registrants), it must be handled with far greater caution. As shown, these disclosures can be exploited by malicious actors.

Website administrators, especially at local government levels, may lack awareness of data subjects' rights under the Personal Data Protection Act (PDPA). Consequently, they may publish sensitive information without consent. This underscores the urgent need for the government to enforce stronger privacy practices, where key measures include:
\begin{itemize}
    \item Educating local web administrators on PDPA compliance.
    \item Providing clear standard guidelines for personal data disclosure.
    \item Mandating the use of anonymization or pseudonymization techniques when personal data must be published for transparency~\cite{10067859}.
\end{itemize}

Additionally, as discussed in Section~\ref{sec:tld}, many exposed documents were found on outdated, decommissioned, or inactive websites. Since such websites may not be actively maintained, exposed personal data could remain accessible for years until the domains expire. This highlights the need for a comprehensive data governance framework to ensure responsible data lifecycle management from collection and storage to decommissioning and deletion.

\subsection{Centralized and Secure Services}
To minimize risks, personal data verification services should be offered via secure, centralized platforms rather than through loosely governed, distributed local websites. A government-backed centralized service with mandatory access controls can ensure that personal data is only accessible to authorized users. This approach would reduce the need for each local agency to publish personal information and would offer a safer, more reliable digital service model aligned with national e-government initiatives.

\subsection{Search Engine Indexing Controls}
As outlined in Section~\ref{sec:searchengine}, search engines index websites based on content, making personal information retrievable via targeted queries. As a last-resort mitigation strategy, web administrators can use the \texttt{noindex} directive to prevent pages from being indexed by search engines~\cite{noindex}. While this does not block direct access to exposed content, it increases the difficulty for malicious actors to locate personal data via search engines, buying valuable time for data owners or agencies to remediate exposure.

\subsection{Legal Accountability and Transparency}
The government must uphold transparency and enforce legal accountability when personal data is leaked. Section~\ref{sec:cybercrime_data} details several previous incidents that should have led to investigations and clear consequences under PDPA. Enforcing these laws uniformly across both government and private sectors will incentivize organizations to adopt better data protection practices, as seen following GDPR enforcement in the European Union~\cite{gdpr_fines}.

\subsection{Proactive Monitoring}
While strong policies and governance frameworks are essential for long-term change, proactive monitoring is crucial to detect and mitigate leaks in real time.

\textbf{Search Engine Monitoring.} As demonstrated in this study, search engines play a central role in exposing personal data. Stakeholders should proactively monitor search engine indexes using specialized tools and keyword tracking. In cases of exposure, search engines can be contacted to request removal of indexed results or cached content~\cite{remove_results, remove_all}. This serves as a critical first response measure, allowing agencies to initiate takedown procedures with website owners before further harm occurs.

\textbf{Threat Intelligence Monitoring.} As noted in Section~\ref{sec:cybercrime_data}, exposed personal data often appears on underground forums and dark web marketplaces. Monitoring these platforms is vital for early detection. Government agencies and security teams should invest in threat intelligence services capable of scanning and analyzing these sources, enabling timely interventions before data is weaponized for fraud, phishing, or identity theft.
\section{Ethics and Responsible Disclosure}
\label{sec:ethics}



Given the sensitive nature of the National ID numbers and personal sensitive data identified in this study, our methodology adheres to the ethical principles outlined in the \textit{Menlo Report}~\cite{menlo_report}, focusing on public interest and harm minimization.

\subsection{Ethical Justification and Proportionality}
The collection of this data is justified by the significant public interest in identifying systemic vulnerabilities in Thailand's digital infrastructure. The scale of the collection (over \note{1.2 million} records) was necessary to demonstrate the proportionality of the threat; a smaller sample size would not have sufficiently illustrated the systemic risk posed by search engine indexing. This study did not involve interaction with human subjects; rather, it analyzed publicly available, albeit sensitive, metadata to identify security failures in government and public sectors.

\subsection{Harm Mitigation and Data Handling}
Our methodology prioritizes the principles of \textit{Beneficence} (minimizing harm while maximizing public benefit) and \textit{Respect for Persons}:

\textbf{Harm Minimization via Air-Gapped Storage:} In accordance with \textit{Menlo Report} guidance on data protection, all retrieved documents and extracted National ID numbers are stored in a secure, offline environment. This ``security-first'' storage protocol ensures that collected PII remains inaccessible to external networks, effectively eliminating the risk of a secondary breach during the research lifecycle.

\textbf{Respect for Autonomy through Anonymization:} While the scale of the exposed data rendered individual informed consent infeasible, we uphold the privacy and dignity of the affected citizens through rigorous data hygiene. We are committed to ensuring that no PII is disclosed to third parties. Only fully anonymized metadata, such as registered domain names and TLD distributions, will be shared, and exclusively for the purposes of advancing security research and mitigating future leakages. No raw personal data will be disclosed under any circumstances.

\textbf{Systemic Evidence Base:} The collection of \note{1.2 million} records was conducted not for individual profiling, but to establish the scale of a systemic national vulnerability. The public interest in identifying and remediating a flaw that affects nearly 2\% of the Thai population outweighs the transient risk of a controlled, academic analysis. Our findings serve as a critical evidence base for the development of automated monitoring systems and the enforcement of Thailand's PDPA.

\subsection{Responsible Disclosure and Takedown Rationale}
We followed a responsible disclosure process rather than an immediate public takedown to ensure systemic remediation. Immediate notification to thousands of individual domain owners was deemed impractical and potentially harmful, as it might alert malicious actors to the vulnerability before a systemic fix was in place. Instead, we have prioritized:

\textbf{Institutional Notification:} We are in the process of coordinating with the relevant Thai government authorities to provide them with the comprehensive list of vulnerable URLs for coordinated remediation.

\textbf{Strategic Mitigation:} By providing these insights to the government sector, we facilitate the development of automated monitoring systems that can prevent future indexing of sensitive files and information, addressing the root cause.
\section{Related Work}
\label{sec:related}

\subsection{Public Sector and Government Data Risks}
Our study reveals that personal data exposure from Thai government websites is not an isolated issue, but part of a broader global challenge tied to digital transformation. As governments worldwide adopt e-Government (e-gov) initiatives to improve efficiency and service delivery~\cite{egov}, they also collect and process vast amounts of personal data. Without strong safeguards, these systems can inadvertently introduce new risks, making them susceptible to data breaches and privacy violations.
Several studies have examined the implications of personal data exposure in the public sector. For instance, Zulfiani (2023) reviewed data leakage incidents in Indonesia and emphasized the urgent need to enhance personal data protection~\cite{zulfiani2021prevention}. Similarly, India’s Aadhaar identity system, which stores biometric and demographic data of over a billion residents, has faced repeated scrutiny over data leaks~\cite{vijay2024survey,SADHYA2024103782}. These cases underscore the challenges of maintaining data security in centralized government systems, which parallels our findings on National ID number exposure in Thailand. At a regional level, Chaipipat et al.~\cite{chaipipat2019asean} examine the fragmented legal frameworks for data protection across ASEAN, highlighting the absence of a unified, enforceable privacy regime. 

The COVID-19 pandemic further amplified privacy concerns globally. Governments collected massive volumes of personal data for vaccination tracking, health monitoring, and contact tracing often under urgent timelines and minimal oversight. Reports from China~\cite{info:doi/10.2196/51219}, Indonesia~\cite{andani2021analysis, ravizki2022criminal}, India~\cite{Malhotrap1407}, and South Korea~\cite{10.3389/fpubh.2020.00305} highlight how emergency responses led to unintended data leaks, public exposure of health records and third-party data sharing. Thailand also faced similar incidents, including the exposure of personal data through vaccination registration application~\cite{9near}, and public releases of recipient lists for COVID-19 financial relief, as analyzed in Section~\ref{sec:eval}.

\subsection{Thailand and Personal Data Privacy}
Thailand’s national development plan, ``Thailand 4.0", envisions a digitally connected society that leverages big data and advanced technologies under the banner of ``Digital for All"~\cite{th_40}. This vision necessarily involves the digitization of personal information across numerous sectors. However, challenges persist in balancing technological advancement with robust privacy protections.


Recent research by Kulrujiphat and Wuttidittachotti (2024) found that public trust in Thailand's e-Government services remains moderate, with concerns over personal data breaches cited as the top issue~\cite{10784415}. These findings reinforce our call in Section~\ref{sec:counter} for stronger infrastructure, centralized access control, and clearer policy enforcement.
Ramasoota and Panichpapiboon highlight key barriers to privacy in Thailand, including national security narratives, weak information practices, and overuse of cybercrime laws for surveillance. Despite the PDPA, these issues continue to erode public trust and legal effectiveness~\cite{th_privacy}.

Compliance with the PDPA also remains a significant hurdle. Chatsuwan et al. (2023) surveyed Thai SMEs and found widespread deficiencies in their privacy practices, including inadequate privacy policies and limited understanding of PDPA~\cite{chatsuwan2023personal}. These findings reflect a broader gap between regulatory intent and implementation, a gap that our study underscores through real-world evidence of data exposure.

\subsection{Search Engine and Personal Data Exposure}
Search engines have long been recognized as powerful tools for uncovering inadvertently exposed documents containing sensitive personal data~\cite{10.1145/1242572.1242652, khan2019privacy, Zimmer_2008, 8563040}. 
Our study confirms this phenomenon within the Thai web ecosystem. We show that National ID numbers and other personal details are retrievable via carefully crafted search queries for personal data fields. These findings align with previous research that explored search engine leakage across various platforms.

For example, a case study involving Scribd, a public document-sharing platform, revealed the exposure of millions of personal records including passport numbers, birth certificates, and phone numbers by using targeted keyword searches refined with Boolean operators~\cite{10823567}. Similarly, data leaks have been identified through search queries within social networking platforms~\cite{8563040, onete2020study}.


\section{Conclusion}
\label{sec:conclusion}

The findings of this study underscore the critical risks associated with the online exposure of personal data, particularly the Thai National ID Number, which serves as a fundamental component of identity verification in both governmental and commercial transactions. Over the course of a three-month data collection period, our research identified 1.2 million exposed National ID numbers, affecting nearly 2\% of Thailand’s population. This significant data exposure raises serious concerns regarding identity theft, financial fraud, and other forms of cybercrime, emphasizing the need for stronger data protection measures.

A key discovery of this study is that the majority of these exposures originate from government sector websites, with domains under the .go.th TLD accounting for 58.8\% of all leaked National ID numbers. Further analysis revealed that even non-government TLDs (.com, .org, .in.th) were often linked to government-affiliated entities, particularly local administrative bodies, and academic institutions. Additionally, sensitive personal details including phone numbers, addresses, financial data, and even health-related information, were found to be publicly accessible, increasing the potential for malicious exploitation.

The implications of this study call for immediate action from the Thai government and relevant stakeholders to strengthen data governance frameworks, enhance cybersecurity protocols, and enforce stricter compliance with the PDPA. Government agencies, in particular, must implement robust data protection measures to prevent unauthorized disclosures. Furthermore, increased public awareness and education on digital security best practices are essential to reducing the likelihood of identity fraud and other cyber threats.

Ultimately, this research serves as a call to action for policymakers, regulatory bodies, and cybersecurity professionals to collaborate in addressing Thailand’s data privacy challenges. By implementing proactive measures such as automated monitoring systems, improved access controls, and stronger legal enforcement, the risk of personal data breaches can be significantly reduced, safeguarding the privacy and security of Thai citizens in an increasingly digital world.

\section{Acknowledgment}
This research was made possible through the generous support of ACinfotec Co., Ltd. under grant contract ACC 036/2567. We extend our sincere gratitude for their invaluable contributions and continued commitment to advancing cybersecurity and data protection research.

\balance
\bibliographystyle{spmpsci}  
\bibliography{main}

\end{document}